# Dynamical Triangulations, a Gateway to Quantum Gravity ?

J. Ambjørn , J. Jurkiewicz[1] and Y. Watabiki

The Niels Bohr Institute
Blegdamsvej 17, DK-2100 Copenhagen Ø, Denmark

## Abstract

We show how it is possible to formulate Euclidean two-dimensional quantum gravity as the scaling limit of an ordinary statistical system by means of dynamical triangulations, which can be viewed as a discretization in the space of equivalence classes of metrics. Scaling relations exist and the critical exponents have simple geometric interpretations. Hartle-Hawkings wave functionals as well as reparametrization invariant correlation functions which depend on the geodesic distance can be calculated. The discretized approach makes sense even in higher dimensional space-time. Although analytic solutions are still missing in the higher dimensional case, numerical studies reveal an interesting structure and allow the identification of a fixed point where we can hope to define a genuine non-perturbative theory of four-dimensional quantum gravity.

[1]Permanent address: Inst. of Phys., Jagellonian University., ul. Reymonta 4, PL-30 059, Kraków 16, Poland

# 1 Introduction

In $d$-dimensional quantum gravity the basic observables are transition amplitudes between $d$-1–dimensional manifolds as well as correlation functions between "objects", e.g. scalar fields, separated by a geodesic distance $r$. We will discuss in detail why the last statement makes sense even if quantum gravity deals with dynamics of fluctuating metrics. Presently no theory of four-dimensional quantum gravity exists and it is not clear if one should attempt to formulate such a theory entirely in field theoretical terms or view it as being embedded in a larger theory (like string theory). As long as the question is not settled we should explore both possibilities. In this article we will review an approach based on conventional field theoretical methods, as they are known from the interplay between field theory and the theory of critical phenomena. The basic idea is to perform a discretization of the theory and in this way make it well defined. In the parameter space of the discretized theory we search for critical points where the "lattice spacing" can be taken to zero and contact made to continuum physics. The discretization is called "dynamical triangulation" and consists of a replacement of the functional integration over all equivalence classes of metrics with a summation over all abstract triangulations of the given manifold. The "lattice" spacing is the length of the links in the triangulations. It is of utmost importance that the discretization can be viewed as a discretization directly in the space of equivalence classes of metrics and not in parameter-space.

We will show that it is possible to formulate Euclidean two-dimensional quantum gravity as the scaling limit of an ordinary statistical system by means of dynamical triangulations. Further, scaling relations can be derived and the critical exponents have simple geometric interpretations. Of course two-dimensional quantum gravity is much simpler than four-dimensional quantum gravity. Nevertheless the study of this theory is very useful since the observables defined above can be calculated analytically. In addition it is possible to test numerical methods, often very useful in the study of critical phenomena, and verify that they work well in two-dimensional quantum gravity. Dynamical triangulations can be used to define a theory of gravity in four dimensions too. In four dimensions we can not solve the model analytically, but it is possible to use numerical methods to get a rather complete picture of the phase diagram of the discretized theory.

# 2 Definition of two-dimensional gravity

## 2.1 Continuum notation

The quantum partition function of two-dimensional quantum gravity can formally be written as

$$Z(\Lambda, G) = \sum_{\mathcal{M} \in \text{Top}} \int_{\mathcal{M}} \frac{\mathcal{D} g_{ab}(\xi)}{\text{Vol(diff)}} \, e^{-S_{\text{eh}}[g;\Lambda,G]} \int \mathcal{D}_g \phi(\xi) \, e^{-S_{\text{m}}[\phi,g]}. \tag{2.1}$$

where $S_{\text{eh}}[g;\Lambda,G]$ is the classical action

$$S_{\text{eh}}(g;\Lambda,G) = \int d^2\xi \sqrt{g} \left[ \Lambda - \frac{1}{4\pi G} \mathcal{R} \right], \tag{2.2}$$

while $S_{\text{m}}$ is the action for some matter fields coupled to gravity. In a full theory of gravity we might have to sum over topologies but it is not yet clarified how to do that except in a perturbative expansion in the genus of the manifold. In the following we will therefore restrict the topology to be spherical, but allow for boundaries on the manifold, i.e. we consider spheres with a number of "holes". For fixed topology the two-dimensional curvature term will play no role since it is a topological invariant and we will ignore it, as well as the matter term, since we will concentrate on the geometric aspects of pure gravity.

Let us write the formal expressions for the objects of interest in pure two-dimensional quantum gravity. The first are generalized Hartle-Hawking wave functionals, i.e. the amplitudes

$$W(L_1, \ldots, L_n) = \int_{S^2_{L_1,\ldots,L_n}} \frac{\mathcal{D}g_{ab}(\xi)}{\text{Vol}(\text{diff})} \, e^{-S[g;\Lambda]} \tag{2.3}$$

for $n$ disconnected one-dimensional universes of prescribed length $L_i$ $i = 1, \ldots, n$. The second class of observables refers more directly to the metric structure of two-dimensional quantum gravity. We can ask about reparametrization invariant correlation functions. The simplest example is:

$$G(R; \Lambda) = \int_{S^2} \frac{\mathcal{D}g_{ab}}{\text{Vol}(\text{diff})} e^{-S[g;\Lambda]} \int\int d^2\xi \sqrt{g(\xi)} d^2\xi' \sqrt{g(\xi')} \, \delta(d_g(\xi,\xi') - R) \tag{2.4}$$

where $d_g(\xi, \xi')$ denotes the geodesic distance with respect to the metric $g_{ab}(\xi)$ and $\Lambda$ is the cosmological constant. It is possible to define a whole set of such reparametrization invariant correlation functions of the geodesic distance by multiplying the densities $\sqrt{g(\xi)} d^2\xi$ and $\sqrt{g(\xi')} d^2\xi'$ in (2.4) by invariant tensors like $\mathcal{R}(\xi)$ and $\mathcal{R}(\xi')$. Note that with the definition (2.4) geodesic distance becomes a meaningful concept even in quantum gravity, despite the fact that we integrate over all metrics.

In certain situations it is convenient to consider universes with a fixed volume $V$. Corresponding to (2.1) we can write

$$Z[V] = \int_{S^2} \frac{\mathcal{D}g_{\mu\nu}}{\text{Vol}(\text{diff})} \, \delta(\int d^2\xi \sqrt{g} - V), \tag{2.5}$$

and in this class of universes the analogue of (2.4) is

$$G(R; V) = \int_{S^2} \frac{\mathcal{D}g_{\mu\nu}}{\text{Vol}(\text{diff})} \int\int d^2\xi \sqrt{g(\xi)} d^2\xi' \sqrt{g(\xi')} \, \delta(d_g(\xi,\xi') - R) \delta(\int d^2\xi \sqrt{g(\xi)} - V). \tag{2.6}$$

It follows directly from the definitions that

$$Z(\Lambda) = \int_0^\infty dV \, e^{-\Lambda V} Z(V), \qquad Z(V) = \int_{c-i\infty}^{c+i\infty} \frac{d\Lambda}{2\pi i} e^{\Lambda V} Z(\Lambda). \tag{2.7}$$

and similarly

$$G(R; \Lambda) = \int_0^\infty dV \, e^{-\Lambda V} G(R; V), \qquad G(R; V) = \int_{c-i\infty}^{c+i\infty} \frac{d\Lambda}{2\pi i} e^{\Lambda V} G(R; \Lambda). \tag{2.8}$$

Note that the short distance behavior of $G(R; V)$ is related to the (internal) *Hausdorff dimension* $d_h$ of the ensemble of manifolds defined by the partition function $Z[\Lambda]$ since it is proportional to the average volume in a "spherical shell" a distance $R$ from an arbitrarily chosen point $\xi$ as long as $R \ll V^{1/d_h}$:

$$G(R; V) \sim R^{d_h - 1} \quad \text{for} \quad R \ll V^{1/d_h}. \tag{2.9}$$

## 2.2 Discretized notation

In two dimensions we have two local invariants, the area element $d^2\xi\sqrt{g}$ and the local curvature $\mathcal{R}(\xi)$. To discretize our manifold we use as building blocks equilateral triangles with side length $a$, i.e. area $\sqrt{3}a^2/4$. We can form an ensemble of piecewise linear manifolds by gluing together the building blocks in all possible ways compatible with the given topology. In this way the order of a given vertex, i.e. the number of triangles which share the vertex, is not fixed. This degree of freedom matches perfectly with the reparametrization invariant degree of freedom given by the local curvature $\mathcal{R}(\xi)$. In fact Regge calculus is precisely a prescription which assigns curvature to piecewise linear manifolds in a way which agrees with the natural concept of parallel transportation on such manifolds. According to Regge calculus [1] the curvature on two-dimensional piecewise linear manifolds lives on the vertices and to vertex $i$ one assigns

$$\mathcal{R}_i\, dA_i = 2(2\pi - \sum_{t \ni i} \Delta\theta_t(i)) \equiv 2\varepsilon_i. \tag{2.10}$$

The summation is over all triangles $t$ which share vertex $i$, and $\Delta\theta_t(i)$ is the angle of triangle $t$ with vertex $i$. If the sum of the angles is $2\pi$, the neighborhood of vertex $i$ is flat. The difference from $2\pi$ is called the *deficit angle*. $dA_i$ stands for a local area element we can assign to the vertex, i.e. it is intended to be $d^2\xi\sqrt{g(\xi)}$ in the neighborhood of vertex $i$. If $A_t$ denotes the area of triangle $t$ it is natural to define

$$dA_i = \frac{1}{3}\sum_{t \ni i} A_t \tag{2.11}$$

if we share the area of a triangle equally between its vertices. With these definitions we have for a closed piecewise linear manifold:

$$\sum_i \mathcal{R}_i\, dA_i = 4\pi\chi, \qquad \sum_i dA_i = \text{area}, \tag{2.12}$$

where $\chi$ denotes the Euler characteristic of the manifold. In our case, where all triangles are equilateral these formulas simplify a lot. If $n_i$ is the order of vertex $i$ and $N_T$ the total number of triangles we have:

$$\mathcal{R}_i\, dA_i = \frac{2\pi}{3}(6 - n_i), \qquad \sum_i dA_i = \frac{\sqrt{3}}{4}a^2 N_T. \tag{2.13}$$

In the following it is convenient to rescale $a$ such that $\sum_i dA_i = a^2 N_T$.

The discretized version of the quantum partition function will be [2, 3, 4]:

$$Z[\mu] = \sum_{T \in \mathcal{T}} \frac{1}{C_T}\, e^{-\mu N_T a^2} \tag{2.14}$$

Here the summation is over a suitable class of (abstract) triangulations $\mathcal{T}$ and $C_T$ denotes the symmetry factor of the triangulation $T$. This factor is present for closed surfaces for the same reason as the special symmetry factors for Feynman vacuum diagrams. $C_T$ is equal to the order of the automorphism group of the graph $T$. In the following we will usually choose $a = 1$, $\sum_i dA_i = N_T$ unless explicitly mentioned otherwise.

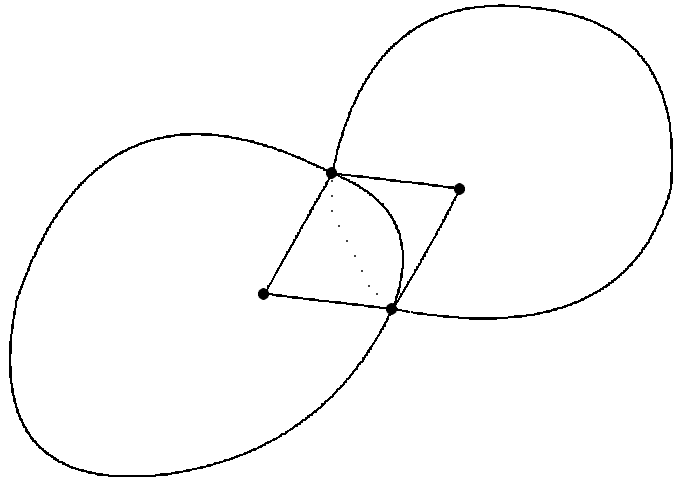

Figure 1: A triangulation with a two-loop, the associated two vertices and the two triangles adjacent to one of the links in the two-loop.

Until now we have not specified precisely which class of triangulations to use, but just stated that we glued the equilateral triangles together to form (closed) manifolds of spherical topology. If we blindly glue $N$ triangles together along the links in order to form a closed manifold we can create somewhat pathological situations where there are closed two-loops (or even one-loops) on the surface even if the building blocks are triangles (see fig. 1). Strictly speaking such graphs cannot be identified with a simplicial manifold according to the standard mathematical definition since the simplical neighborhood of a vertex is not necessarily a disk. Whether we allow such two-loops (or one-loops) or not should not be important for a continuum limit since these are just structures at the cut-off scale. It has been verified *a posteriori* that it does not matter precisely which class of triangulations we use, as long as they allow an identification of the topology of the given piecewise linear manifold. We can write (2.14) as

$$Z(\mu) = \sum_{N=1}^{\infty} e^{-\mu N} Z(N), \quad Z(N) = \sum_{T \in \mathcal{T}_N} \frac{1}{C_T}. \qquad (2.15)$$

where $\mathcal{T}_N$ denotes the triangulations made of $N$ triangles. We denote $Z(\mu)$ the *grand canonical* partition function and $Z(N)$ the *canonical* partition function since we can view $\mu$ as the chemical potential for creating additional triangles. $Z(N)$ has the interpretation as the number $\mathcal{N}(N)$ of triangulations in $\mathcal{T}_N$. This number is exponentially bounded [5]:

$$Z(N) = \mathcal{N}(N) = e^{\mu_c N} N^{\gamma-3}(1 + O(1/N)) \qquad (2.16)$$

As a consequence $Z(\mu)$ is well defined for $\mu > \mu_c$ and divergent for $\mu < \mu_c$. The point $\mu_c$ is the *critical point* and we should define the continuum limit as $\mu \to \mu_c$. In this limit the large $N$ part of the distribution will dominate in the sense that we can write:

$$Z(\mu) \sim (\mu - \mu_c)^{2-\gamma} + \text{less singular terms}. \qquad (2.17)$$

While the exponent $\gamma$ has the interpretation as power correction to the exponential growing number of triangulations and thereby determines the leading singularity of partition function, it also determines the part of the fractal structure related to the tendency to branch into so called *baby universes*. Let us first define a baby universe as part of the piecewise linear surface separated from the rest by a minimal "bottleneck". The precise nature of such a bottleneck depends on the class of triangulations we use. If we consider only regular triangulations such a bottleneck will be a loop of three links. If we cut our piecewise linear manifold along this loop we will separate it in two. Usually this separation will be trivial in the sense that the loop will be just the boundary of a triangle and all we have done is to separate this triangle from the rest of the triangulation. However, in case the loop is not the boundary of a triangle belonging to the triangulation cutting along the loop will produce a separation into two non-trivial parts of the triangulation. The smaller part is called a "minimal bottleneck baby universe", abbreviated "minbu" [6], the larger part the "parent".

A moment of reflection will convince the reader that the average number of baby universes of volume $V$ in the ensemble of triangulations of total volume $N$ will be given by

$$\langle \mathcal{N}(V) \rangle_N \approx \frac{3!}{Z(N)} V Z(V) \, (N-V) Z(N-V) \tag{2.18}$$

where 3! is the number of ways one can glue the two boundaries of the minbu and the parent together. The additional factors $V$ and $N - V$ reflect the fact that the minbu and the parent both have a minimal boundary i.e. for a generic large surface of volume $V$ and no accidental symmetry factors there will be $V$ such manifolds for each closed manifold since the boundary can be placed at any of the $V$ triangles. If we *assume* that the canonical partition function is given by (2.16) we get:

$$\langle \mathcal{N}(V) \rangle_N \sim N \, [V(1 - V/N)]^{\gamma - 2} . \tag{2.19}$$

which has the obvious interpretation that the probability density for branching to a baby universe of volume $V$ is $V^{\gamma-2}(1 - V/N)^{\gamma-2}$.

## 2.3  Scaling relations

In addition to the entropy exponent $\gamma$ we can introduce the critical exponents $\nu$ and $\eta$ known from the theory of critical phenomena: $\nu$ is the exponent for the inverse mass or the correlation length and $\eta$ is the anomalous scaling exponent. Further, $\gamma$ can be given an interpretation as a susceptibility exponent.

In order to define these exponents [7, 8, 26] let us first consider the discretized analogy of the continuum two-point function $G(R; \Lambda)$ given by (2.4). For each surface, built of equilateral triangles, we have by Regge's prescription a metric. This means that we can define the geodesic distance between any two points on the piecewise linear manifold. Here we use a simplified definition instead. The geodesic distance between two triangles is defined by considering the path connecting centers of triangles. A geodesic path is one where the length is minimal, and the length is usually counted in units of the lattice spacing $a$. In the same way we can define the geodesic distance between two links as the shortest path along a sequence of neighboring triangles which connects the two links and we can define the geodesic distance between a link and a set of links as the shortest geodesic distance to a member of the set. Let us now consider the sub-ensemble of surfaces where the geodesic distance between two links $l_1$ and $l_2$ is fixed to be $r$ (which in units of the lattice spacing $a$ is an integer). Denote this ensemble of surfaces by $\mathcal{T}(2,r)$. We define the two-point function by:

$$G_\mu(r) = \sum_{N=1}^{\infty} e^{-\mu N} \sum_{T \in \mathcal{T}_N(2,r,N)} 1. \qquad (2.20)$$

If we have two marked links the combinatorial factor $C_T = 1$. $G_\mu(r)$ is the discretized version of the volume-volume correlator[2] $G(R;\Lambda)$.

We have the following theorem:

**Theorem:** The two-point function falls off exponentially for $r \to \infty$.

$$\lim_{r \to \infty} \frac{-\log G_\mu(r)}{r} = m(\mu) \geq 0. \qquad (2.21)$$

In addition the mass $m(\mu) > 0$ for $\mu > \mu_c$ and $m(\mu)$ is a decreasing function of $\mu$.

This trivial but important relation follows from the fact that

$$G_\mu(r_1 + r_2) \geq G_\mu(r_1) G_\mu(r_2), \qquad (2.22)$$

simply because each term on the rhs of eq. (2.22) can be given an interpretation as a term belonging to the lhs of eq. (2.22). This is illustrated in fig. 2. Eq. (2.22) shows that $-\log G_\mu(r)$ is sub-additive and this ensures the existence of the limit (2.21). In addition $m(\mu) \geq 0$ because $G_\mu(r)$ is a decreasing function of $r$. Again this follows from general arguments which allow us to bound the number of triangulations with $N$ triangles and two marked links separated by a distance $r$ in terms of the number of triangulations with $N$ triangles and two marked links separated by a distance $r' < r$. The same kind of arguments leads to the conclusion that $m'(\mu) > 0$ for $\mu > \mu_c$, i.e. $m(\mu)$ is a decreasing function for $\mu \to \mu_c$.

Consequently we expect the following generic behavior of the two-point function:

$$G_\mu(r) \sim e^{-m(\mu)r}, \qquad r \gg 1/m(\mu); \qquad (2.23)$$

$$G_\mu(r) \sim r^{1-\eta}, \qquad 1 \ll r \ll 1/m(\mu); \qquad (2.24)$$

$$\chi(\mu) \equiv \sum_r G_\mu(r) \sim (\mu - \mu_c)^{-\gamma} + \text{less singular terms}. \qquad (2.25)$$

We use the exponent $r^{1-\eta}$ for the short distance behavior in order to be in accordance with the general notation:

$$G(r) \sim r^{d-1} \frac{1}{r^{d-2+\eta}}, \qquad (2.26)$$

where the first factor is due to the angular average over a spherical shell of radius $r$. Note that (2.25) gives an alternative interpretation of $\gamma$ as a *susceptibility exponent*. Recall that for a spin system the susceptibility is the second derivative of the free energy with respect to the external magnetic field and it has the alternative interpretation as the integral of the spin-spin correlation function. The same is true here: By (2.15) and (2.17) it follows that the second derivative of $Z(\mu)$ with respect to $\mu$ has a leading singularity $(\mu - \mu_c)^{-\gamma}$, while eq. (2.25) provides us with the alternative interpretation as the integral of the volume-volume correlator. $\chi(\mu) \sim Z''(\mu)$ since both can be viewed as the summation over all triangulations with two marked triangles or links.

---

[2]It is possible to associate an invariant volume $2a^2/3$ with each link since it has two adjacent triangles and each triangle contains three links. Alternatively we could have defined the two-point function using directly marked triangles instead of marked links. Since we are later going to consider loop-loop correlation functions the first definition is technically more convenient in two dimensions.

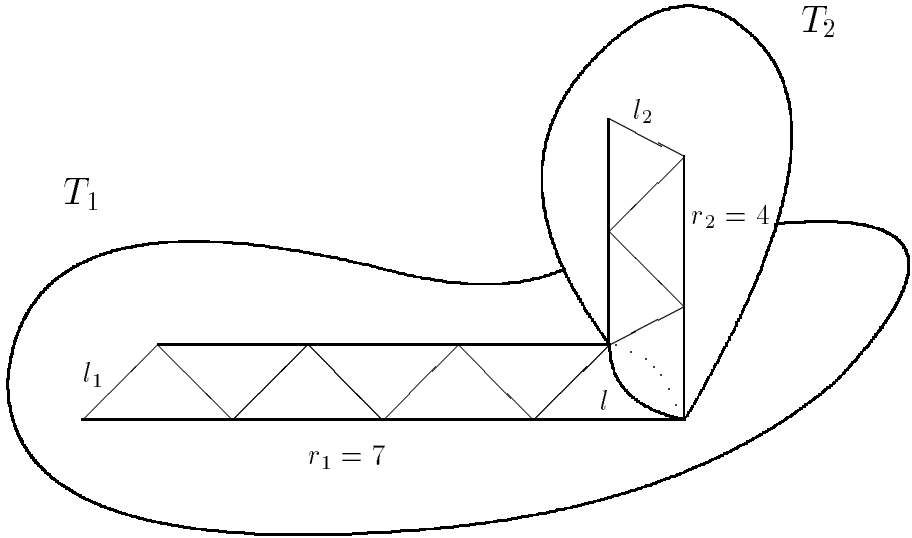

Figure 2: The inequality (2.22). Two triangulations with marked links separated by distances $r_1$ and $r_2$ can be glued together to a triangulation where the marked links have a distance $r_1 + r_2$ but the same number of triangles by cutting open a marked link in each of the triangulations to a 2-loop boundary and gluing together the two boundaries.

Although we have proven above that the mass decreases as $\mu \to \mu_c$ the general arguments do not allow us to conclude that $m(\mu)$ scales to zero at the critical point $\mu_c$. Let us add this as an *assumption* (we will later verify that it is true in two-dimensional quantum gravity):

$$m(\mu) \sim (\mu - \mu_c)^\nu, \tag{2.27}$$

and explore the consequences. In order to make the scaling relations more transparent let us for a moment reintroduce the dimensionful lattice spacing $a$ and define the *renormalized* cosmological constant $\Lambda$ in terms of the *bare* cosmological constant $\mu$ by an additive renormalization:

$$\mu = \mu_c + \Lambda a^2, \quad \text{i.e.} \quad a(\mu) \sim (\mu - \mu_c)^{1/2}. \tag{2.28}$$

If we want a scaling limit where the concept of a mass survives we *have* to introduce a scaling like:

$$m(\mu) = M\,[a(\mu)]^{2\nu}, \quad \text{i.e} \quad M = c\Lambda^\nu, \tag{2.29}$$

$$R = r\,[a(\mu)]^{2\nu}, \tag{2.30}$$

where $c$ is a constant of order one. In this way the scaling limit is one where the "bare" mass $m(\mu)$ scales to zero as the lattice spacing vanishes, while the "renormalized" mass $M$ is kept fixed. Similar remarks apply for the "bare" geodesic distance $r$ and the "renormalized" geodesic distance, only will $r$ go to infinity for $a(\mu) \to 0$. The long distance behavior of the two-point function will be

$$G_\mu(r) \sim e^{-MR} \quad \text{for} \quad \mu \to \mu_c, \quad m(\mu)\, r \gg 1. \tag{2.31}$$

The relation between the continuum propagator $G(R; \Lambda)$ and $G_\mu(r)$ is unambiguously fixed by (2.23)-(2.25):

$$G(R; \Lambda) = \lim_{\mu \to \mu_c} [a(\mu)^{2\nu}]^{1-\eta}\, G_\mu(r), \quad m(\mu)\, r = MR. \tag{2.32}$$

In addition the only possibility for this limit to exist is that the long distance behavior of $G_\mu(r)$ is given by

$$G_\mu(r) = f(m(\mu)r)\, [a(\mu)^{2\nu}]^{\eta-1}\, e^{-m(\mu)r}, \tag{2.33}$$

where $f(x)$ (which we ignored in (2.23)) is subleading compared to the exponential function, i.e.

$$G(R; \Lambda) \sim R^{1-\eta}, \ c\Lambda^\nu R \ll 1, \quad G(R; \Lambda) \sim f(c\Lambda^\nu R)\, \Lambda^{\nu(\eta-1)}\, e^{-c\Lambda^\nu R}, \ c\Lambda^\nu R \gg 1. \tag{2.34}$$

We shall verify that we indeed have the correct asymptotic behavior (2.32) (with $f(x) = $ const. and $\nu = 1/4$) in pure two-dimensional quantum gravity.

In the scaling limit (2.31) the exponent $\nu$ has the simple geometric interpretation as the inverse of the *internal* Hausdorff dimension of the ensemble of piecewise linear surfaces $\mathcal{T}(2, r)$. Let $\langle \cdot \rangle_r$ denote the average with respect to the ensemble $\mathcal{T}(2, r)$. We define the internal Hausdorff dimension $d_h$ by

$$\langle N \rangle_r \sim r^{d_h} \quad \text{for} \quad r \to \infty, \quad m(\mu)\, r = \text{const}. \tag{2.35}$$

From (2.23) we have

$$\langle N \rangle_r = -\frac{\partial \log G_\mu(r)}{\partial \mu} = \frac{\partial m(\mu)}{\partial \mu}\, r, \tag{2.36}$$

and using (2.27) we get

$$\langle N \rangle_r \sim r^{1/\nu}, \quad \text{i.e.} \ d_h = 1/\nu. \tag{2.37}$$

Another consequence of the above scaling is that we can rewrite (2.36) as:

$$\langle N \rangle_R \sim \frac{MR}{\mu - \mu_c}. \tag{2.38}$$

This relation shows how the number of triangles diverges as $\mu \to \mu_c$, (as we want in order to connect to continuum physics) and in terms of the dimensionful volume $V = Na^2$ we have:

$$\langle V \rangle_R \sim \frac{R}{\Lambda^{1-\nu}}, \quad c\Lambda^\nu R \gg 1. \tag{2.39}$$

This relation simply tells us that the typical universe which for a given cosmological constant $\Lambda$ has points separated by a "continuum" geodesic distance $R$ which is much larger that $1/\Lambda^{1-1/d_h}$ will be a long tube.

It is interesting to give a direct physical interpretation of the short distance behavior of $G_\mu(r)$ as defined by (2.24). In order to do so let us change from the grand canonical ensemble given by (2.20) to the canonical ensemble defined by $\mathcal{T}(2,r,N)$, the class of triangulations with $N$ triangles where two links (or triangles) are marked and separated by a distance $r$. On this ensemble we can define the discretized analogue of $G(R,V)$:

$$G(r,N) = \sum_{T \in \mathcal{T}(2,r,N)} 1, \qquad (2.40)$$

For $r = 0$ we have the following $N$ dependence (for the $r$ dependence see (2.45) below)

$$G(0,N) \sim N^{\gamma-2} e^{\mu_c N}. \qquad (2.41)$$

The partition function $Z(N) \sim N^{\gamma-3} e^{\mu_c N}$ and the one-point function for large $N$ is proportional to $NZ(N)$ since it counts the triangulations with one marked triangle (or link or vertex depending on the precise definition). For $r = 0$ (or just small) there is essentially no difference between the one-point function and $G(0,N)$.

$G(r,N)$ is related to $G_\mu(r)$ by a (discrete) Laplace transform:

$$G_\mu(r) = \sum_N G(r,N) \, e^{-\mu N}. \qquad (2.42)$$

The *long distance behavior* of $G(r,N)$ is determined by the long distance behavior of $G_\mu(r)$. Close to the scaling limit it follows by direct calculation (e.g. a saddle point calculation[3]) that

$$\begin{aligned} G_\mu(r) &\sim e^{-r(\mu-\mu_c)^{1/d_h}} \quad \Rightarrow \\ G(r,N) &\sim e^{-c\left(r^{d_h}/N\right)^{\frac{1}{d_h-1}}} e^{\mu_c N} \quad \text{for} \quad r^{d_h} > N, \end{aligned} \qquad (2.43)$$

where $c = (d_h - 1)/d_h^{d_h/(d_h-1)}$.

On the other hand the *short distance behavior* of $G_\mu(r)$ is determined by the short distance behavior of $G(r,N)$ which is simple. Eqs. (2.35) and (2.37) define the concept of Hausdorff dimension in the grand canonical ensemble. A definition in the canonical ensemble would be: Take $N^{1/d_h} \gg r$ and simply count the volume (here number of triangles) of a "spherical shell" of thickness 1 and radius $r$ from a marked triangle, sum over all triangulations $T_N$ with one marked triangle and divide by the total number of such triangulations. Call this number $\langle n(r) \rangle_N$. The Hausdorff dimension is then defined by

$$\langle n(r) \rangle_N \sim r^{d_h - 1} \quad \text{for} \quad 1 \ll r \ll N^{1/d_h}. \qquad (2.44)$$

It follows from the definitions that we can write

$$\begin{aligned} \langle n(r) \rangle_N &\sim \frac{G(r,N)}{G(0,N)}, \quad \text{i.e} \\ G(r,N) &\sim r^{d_h-1} N^{\gamma-2} e^{\mu_c N} \quad \text{for} \quad 1 \ll r \ll N^{1/d_h}. \end{aligned} \qquad (2.45)$$

---

[3]In addition to the exponentially decaying part of $G(r,N)$ there is also a power correction coming from the quadratic integration in the saddle point approximation. We shall not consider the explicit form of the power correction here.

We can finally calculate the short distance behavior of $G_\mu(r)$ from eq. (2.42). For $\mu \to \mu_c$, we get:

$$G_\mu(r) \sim r^{d_h-1} \sum_N N^{\gamma-2} e^{-c\left(r^{d_h}/N\right)^{\frac{1}{d_h-1}}} \sim r^{\gamma d_h - 1} \qquad (2.46)$$

This shows that

$$\eta = 2 - \gamma d_h, \quad \text{i.e.} \quad \gamma = \nu(2 - \eta), \qquad (2.47)$$

which is *Fishers scaling relation*, which is valid also for quantum gravity. Of course the relation could be derived directly from (2.23)-(2.25), but the above arguments highlight that the anomalous scaling dimension $\eta$ is a function of the two kinds of fractal structures we can define on the ensemble of piecewise linear manifolds: the Hausdorff dimension and the baby universe proliferation probability. In addition the arguments show that the canonical and grand canonical definitions of Hausdorff dimension agree.

A most important remark is that the definitions and the scaling relations above generalize to higher dimensional quantum gravity.

## 2.4 Branched polymers

The model of branched polymers ($BP$) provides us with a simple, but non-trivial example of the above scenario [10] and will play an important role in the following. In a certain way it can be viewed as the lowest dimensional fractal structure and it will appear as the limiting case of higher dimensional gravity theories.

Let us define branched polymers as the sum over all tree graphs (no loops in the graphs) with certain weights given to the graphs according to the following definition of the partition function:

$$Z(\mu) = \sum_{BP} \frac{1}{C_{BP}} \rho(BP) \, e^{-\mu|BP|}, \qquad (2.48)$$

where $|BP|$ is the number of links in a $BP$ and $\mu$ is a chemical potential for the number of links, while

$$\rho(BP) = \prod_{i \in BP} f(n_i), \qquad (2.49)$$

where $i$ denotes a vertex, $n_i$ the number of links joining at vertex $i$ and $f(n_i)$ is non-negative. $f(n_i)$ can be viewed as the unnormalized branching weight for one link branching into $n_i - 1$ links at vertex $i$. Finally $C_{BP}$ is a symmetry factor such that rooted branched polymers, i.e. polymers with the first link marked, is counted only once.

This model can be solved [10, 11]. It has a critical point $\mu_c$ (depending on $f$) and close to the critical point we have:

$$Z''(\mu) \sim (\mu - \mu_c)^{-1/2}, \qquad (2.50)$$

i.e. $\gamma = 1/2$ for branched polymers. On the branched polymers we define the "geodesic distance" between two vertices as the shortest link path, which is unique since we consider tree-graphs. The graphic representation of the two-point function is shown in fig. 3. Had it not been for the ability to branch, the two-point function would simply be

$$G_\mu(r) = e^{-\mu r}. \qquad (2.51)$$

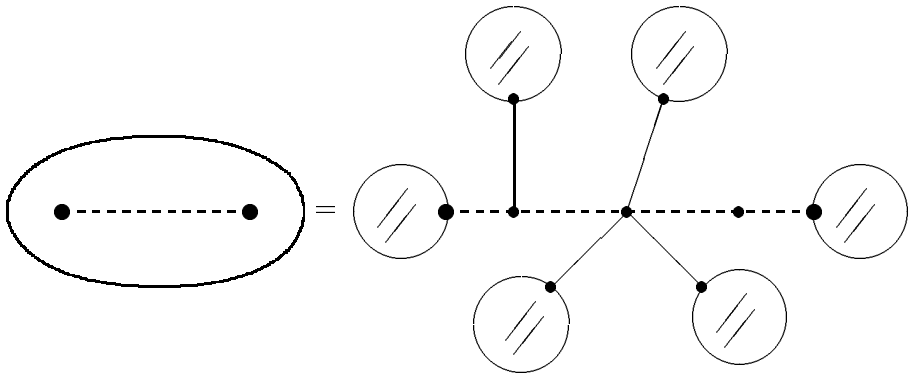

Figure 3: The graphical representation of the two-point function for branched polymers. The dashed line represents the unique shortest path between the two marked vertices. The "blobs" represent the contribution from all rooted polymers branching out from a vertex.

However, the insertion of one-point functions at any vertex leads to a non-analytic coupling constant renormalization and the result is changed to [10]

$$G_\mu(r) = \text{const.}\, e^{-\kappa\, r\sqrt{\mu-\mu_c}} \quad \text{for} \quad \mu - \mu_c \to 0, \tag{2.52}$$

where $\kappa$ is a positive constant depending on $f$. We can now find $G(r, N)$ by an inverse Laplace transform:

$$G(r, N) = \text{const.}\, N^{-3/2} r\, e^{-\kappa^2 r^2/4N}. \tag{2.53}$$

We confirm from this explicit expression that the (internal) Hausdorff dimension of branched polymers is 2 (like a smooth surface !) and that $\gamma = 1/2$ since the prefactor of $G(r, N)$ for small $r$ should be $N^{\gamma-2} r^{d_H-1}$.

## 2.5 Perspectives

We have derived general scaling relations for quantum gravity. In the following we will solve the theory explicitly in two dimensions. It is remarkable that both the Hartle-Hawkings wave-functionals and the volume-volume correlation functions can be found by purely combinatorial methods at the discretized level, after which the scaling limit can be taken unambiguously.

## 3 Matrix models

In this section we will discuss how to construct Hartle-Hawkings wave functionals in the discretized approach. The problem was solved already in 1963 by the mathematician Tutte, who found the generating functionals for the number of planar triangulations with boundaries. We shall rederive here the results using the so called matrix models technique, following the approach in [12, 13].

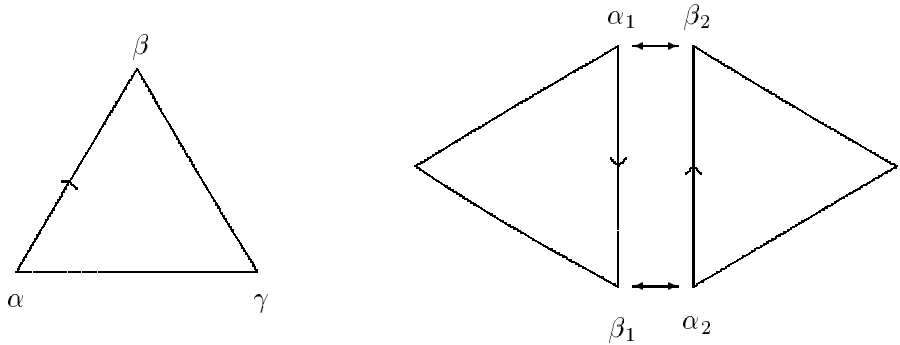

Figure 4: The matrix representation of triangles which converts the gluing along links to a Wick contraction.

## 3.1 The Hartle-Hawking wave functionals

In order to count the number of triangulations we represent the triangles by means of Hermitian matrices: Label the vertices of the $i^{th}$ triangle by abstract indices $\alpha_i, \beta_i, \gamma_i$ and attach a Hermitian matrix $\phi_{\alpha_i \beta_i}$ to the oriented link from $\alpha_i$ to $\beta_i$. In this way we can attach the scalar quantity

$$\phi_{\alpha_i\beta_i}\phi_{\beta_i\gamma_i}\phi_{\gamma_i\alpha_i} = \text{Tr}\,\phi^3 \tag{3.1}$$

to each of the $K$ triangles. The Gaussian integral

$$\int d\phi\, e^{-\frac{1}{2}\text{Tr}\,\phi^2} \frac{1}{K!}\left(\frac{1}{3}\text{Tr}\,\phi^3\right)^K \tag{3.2}$$

where

$$d\phi \equiv \prod_{\alpha \leq \beta} d\Re\phi_{\alpha\beta} \prod_{\alpha < \beta} d\Im\phi_{\alpha\beta} \tag{3.3}$$

can be performed by doing all possible Wick contractions of $\phi$-fields. This corresponds to performing all possible gluings of surfaces of $K$ triangles, the reason being that each Wick contraction in the Gaussian integral glues together two links:

$$\left\langle \phi_{\alpha_i\beta_i}\phi_{\alpha_j\beta_j} \right\rangle \equiv \int d\phi\, e^{-\frac{1}{2}|\phi_{\alpha\beta}|^2} \phi_{\alpha_i\beta_i}\phi_{\alpha_j\beta_j} = \delta_{\alpha_i\beta_j}\delta_{\beta_i\alpha_j}. \tag{3.4}$$

This is illustrated in fig. 4. After all Wick contractions are performed in eq. (3.2) the $K$ triangles have been glued together in all possible ways. The surfaces created in this way will consist of disconnected parts, but we get the connected graphs by taking the logarithm of all graphs. Furthermore we can calculate the contribution from a particular graph constituting a closed surface: we pick up a factor $N$, $N$ being the number of indices, whenever a vertex becomes an internal vertex in the process of gluing together links by Wick contractions. This means that we get a total factor $N^V$, where $V$ is the number of vertices. If we make the substitution

$$\text{Tr}\,\phi^3 \to \frac{g}{\sqrt{N}}\text{Tr}\,\phi^3 \tag{3.5}$$

it is seen that the factors of $N$ for $K$ triangles combine to $N^{V-K/2} = N^\chi$, since the Euler characteristic for a triangulation of $K$ triangles, $L$ links and $V$ vertices is

$$\chi = V - L + K = V - K/2. \tag{3.6}$$

By multiplying each triangle $\text{Tr}\,\phi^3$ by $1/\sqrt{N}$ the weight of a given triangulation only depends on its topology. In addition the sum over all triangulations exponentiates. Collecting this information we can write:

$$Z(\mu, G) = \log \frac{Z(g, N)}{Z(0, N)} \tag{3.7}$$

where $Z(\mu, G)$ is partition function defined in terms of dynamical triangulations and

$$Z(g, N) = \int d\phi \, \exp\left(-\frac{1}{2}\text{Tr}\,\phi^2 + \frac{g}{3\sqrt{N}}\text{Tr}\,\phi^3\right) \tag{3.8}$$

provided we make the identification:

$$\frac{1}{G} = \log N, \qquad \mu = -\log g. \tag{3.9}$$

Eq. (3.8) allows a $1/N^2$ expansion and this expansion is an expansion in topology. Here we will only be interested in spherical topology, i.e. the leading term in the $1/N^2$ expansion.

For the purpose of a general (perturbative) analysis of the matrix integral (3.8) it is convenient to consider the generalization to an arbitrary set of coupling constants $\{g_i\}$:

$$Z(g_1, g_2, \ldots) = \int d\phi \, e^{-N\text{Tr}\,V(\phi)} \tag{3.10}$$

where

$$V(\{g_i\}) = \sum_{n=1}^{\infty} \frac{g_n}{n}\phi^n. \tag{3.11}$$

In eq. (3.11) we have of convenience scaled $\phi \to \sqrt{N}\phi$. In this way the topological nature of the expansion is still preserved: All two-dimensional complexes of Euler characteristic $\chi$ will have a factor $N^\chi$ associated with them. The interpretation of (3.10) is intended to be as before: we have in mind a Gaussian integral around which we expand, i.e. $g_2 > 0$ and $g_n \leq 0$ with the sign convention used in (3.11). The convenience of considering an arbitrary potential is that the general coupling constants $g_n$ act as sources for terms like $\text{Tr}\,\phi^n$, and by differentiating $Z$ with respect to $g_n$ we can calculate expectation values of "observables" like $\text{Tr}\,\phi^n$. $\langle \text{Tr}\,\phi^n/N \rangle$ has the following obvious interpretation: It represents the summation over all "surfaces" which have a $n$-sided polygon with one marked link as boundary. This follows from the gluing procedure realized by Wick contractions of the Gaussian integrals[4]. In a similar way

$$\frac{1}{N^2}\langle \text{Tr}\,\phi^n \text{Tr}\,\phi^m \rangle - \frac{1}{N^2}\langle \text{Tr}\,\phi^n \rangle \langle \text{Tr}\,\phi^m \rangle \tag{3.12}$$

will represent the sum over all connected two-dimensional complexes which connect one boundary consisting of $n$ links with another boundary consisting of $m$ links, *i.e. precisely the discretized version of the Hartle-Hawking wave functionals*. Let us define the generating

---

[4]In case we want unmarked links we should multiply $\langle \text{Tr}\,\phi^n/N \rangle$ with an additional symmetry factor $1/n$.

functional for connected loop correlators. The expectation value of an arbitrary observable is defined by

$$\langle Q(\phi) \rangle \equiv \frac{1}{Z} \int d\phi \, e^{-N \operatorname{Tr} V(\phi)} \, Q(\phi). \tag{3.13}$$

The generating function for $s$-loop correlators, which we will also, somewhat inaccurate, denote the $s$-loop correlator, is defined by

$$W(z_1, \ldots, z_s) \equiv N^{s-2} \sum_{k_1, \ldots, k_s}^{\infty} \frac{\langle \operatorname{Tr} \phi^{k_1} \cdots \operatorname{Tr} \phi^{k_s} \rangle_{conn}}{z^{k_1+1} \cdots z^{k_s+1}} = N^{s-2} \langle \operatorname{Tr} \frac{1}{z_1 - \phi} \cdots \operatorname{Tr} \frac{1}{z_n - \phi} \rangle_{conn}. \tag{3.14}$$

where $conn$ refers to the connected part as defined by (3.12), or its generalization to more correlators. If we introduce the so-called *loop insertion operator* by

$$\frac{d}{dV(z)} \equiv -\sum_{k=1}^{\infty} \frac{k}{z^{k+1}} \frac{d}{dg_k}. \tag{3.15}$$

it follows from the definition (3.14) that

$$W(z_1, \ldots, z_s) = \frac{d}{dV(z_s)} \frac{d}{dV(z_{s-1})} \cdots \frac{d}{dV(z_2)} W(z_1) \tag{3.16}$$

and this equation shows that *if the one-loop operator is known for an arbitrary potential, all multi-loop correlators can be calculated.*

The one-loop correlator is related to the density $\rho(\lambda)$ of eigenvalues defined by the matrix integral as follows:

$$\rho(\lambda) = \langle \sum_{i=1}^{N} \delta(\lambda - \lambda_i) \rangle \tag{3.17}$$

where $\lambda_i$, $i = 1, \ldots, N$ denote the $N$ eigenvalues of the matrix $\phi$. With this definition we have

$$W(z) = \int_{-\infty}^{\infty} d\lambda \, \frac{\rho(\lambda)}{z - \lambda} \tag{3.18}$$

For $N \to \infty$ there exist, as we shall see, consistent solutions where the support of $\rho$ is confined to a finite interval $[c_0, c_1]$ on the real axis. In this case $W(z)$ will be an analytic function in the complex plane, except for a cut at the support of $\rho$ and it follows from Schwartz's reflection principle that

$$2\pi i \rho(\lambda) = \lim_{\varepsilon \to 0} W(\lambda + i\varepsilon) - W(\lambda - i\varepsilon) \tag{3.19}$$

## 3.2 The loop equations

Let us explore the invariance of the matrix integral (3.10) under field redefinitions of the type:

$$\phi \to \phi + \varepsilon \sum_{k=0}^{\infty} \frac{\phi^k}{p^{k+1}} = \phi + \varepsilon \frac{1}{p - \phi}. \tag{3.20}$$

This kind of field redefinitions only make sense if $p$ is chosen on the real axis outside the support of the eigenvalues of $\phi$. As mentioned above we will verify that this scenario is realized for $N \to \infty$. Under the transformation (3.20) the measure and the action change as

$$d\phi \to d\phi \left(1 + \varepsilon \text{Tr} \frac{1}{p - \phi} \text{Tr} \frac{1}{p - \phi}\right) \tag{3.21}$$

$$\text{Tr}\, V(\phi) \to \text{Tr}\, V(\phi) + \varepsilon \text{Tr} \left(\frac{1}{p - \phi} V'(\phi)\right). \tag{3.22}$$

The integral (3.10) will be invariant under a redefinition of the integration variables by eq. (3.20) and the change of measure and action has to cancel to first order in $\varepsilon$. By use of eqs. (3.21) and (3.22) this leads to the following equation:

$$\int d\phi \left\{\left(\text{Tr} \frac{1}{p - \phi}\right)^2 - N \text{Tr} \left(\frac{1}{p - \phi} V'(\phi)\right)\right\} e^{-N \text{Tr}\, V(\phi)} = 0. \tag{3.23}$$

The first term in this equation is by definition

$$N^2 W(p) W(p) + W(p, p). \tag{3.24}$$

The second term in eq. (3.23) can be written as an integral over the the one-loop correlator by means of the density of eigenvalues and we can finally write (3.23) in the standard form, known as the loop equation [14]:

$$\oint_C \frac{d\omega}{2\pi i} \frac{V'(\omega)}{z - \omega} W(\omega) = W(z)^2 + \frac{1}{N^2} W(z, z) \tag{3.25}$$

where $z$ is outside the interval $[c_0, c_1]$ on the real axis and the contour $C$ encloses the cut, but not $z$ (see fig. 5). In eq. (3.25) the $1/N^2$ expansion is manifest and since we are interested in the leading term (spherical topology) we ignore the last term on the rhs of (3.25) and denote the corresponding solution $W_0$.

In case of a Gaussian potential $V(z) = z^2/2$ the eigenvalue density $\rho(\lambda)$ is given by famous Wiener's semicircle law:

$$\rho(\lambda) = \frac{1}{2\pi} \sqrt{(2 - \lambda)(2 + \lambda)} \quad \text{i.e.} \quad W_0(z) = \frac{1}{2} \left(z - \sqrt{(z - 2)(z + 2)}\right), \tag{3.26}$$

where the last equation follows from (3.18).

For a general potential we can find a solution which has essentially the same structure as for the quadratic potential and where the eigenvalue density is given by [15]:

$$\rho(\lambda) \sim M(\lambda) \sqrt{(c_1 - \lambda)(\lambda - c_0)}, \tag{3.27}$$

where $M(\lambda)$ is a polynomial of order $n - 2$ if $V(\lambda)$ is a polynomial of order $n$. We write:

$$W_0(z) = \frac{1}{2} \left(V'(z) - M(z) \sqrt{(z - c_1)(z - c_0)}\right) \tag{3.28}$$

and the requirement that $W_0(z) = O(1/z)$ uniquely determines $M(z)$ since (3.28) allows us to write

$$M(z) = \oint_{C_\infty} \frac{d\omega}{2\pi i} \frac{M(\omega)}{z - \omega} = \oint_{C_\infty} \frac{d\omega}{2\pi i} \frac{V'(\omega)}{z - \omega} \frac{1}{\sqrt{(\omega - c_1)(\omega - c_0)}}, \tag{3.29}$$

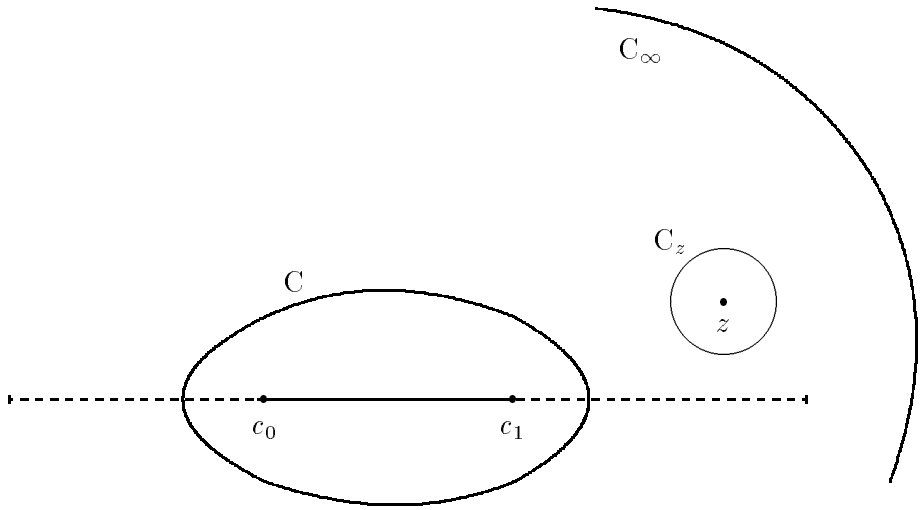

Figure 5: The integration contour $C$ and the cut from $c_0$ to $c_1$. When deforming the contour to infinity we get two contributions: one from the circle $C_\infty$ and one from the circle $C_z$ around the pole $z$.

as the part of the integral which involves $W_0(z)$ vanishes. By expanding the last integrand in powers of $1/(\omega - c_1)$ we can write

$$M(z) = \sum_{k=1}^{\infty} M_k (z-c_1)^{k-1}, \qquad M_k[c_0, c_1; g_i] = \oint_C \frac{d\omega}{2\pi i} \frac{V'(\omega)}{(\omega - c_1)^{k+\frac{1}{2}}(\omega - c_0)^{\frac{1}{2}}}. \qquad (3.30)$$

It is important to note that $M_k = 0$ for $k \geq n$ where $n$ is the order of the polynomial $V(z)$. It follows directly from the integral which defines $M_k$.

$W_0(z)$ can now be expressed in terms of the "moments" $M_k$ [12, 13]:

$$W_0(z) = \frac{1}{2}\left(V'(z) - \sqrt{(z-c_1)(z-c_0)} \sum_{k=1}^{\infty} M_k[c_0, c_1; g_i](z-c_0)^{k-1}\right). \qquad (3.31)$$

The endpoints $x$ and $y$ of the cut are determined selfconsistently by the following boundary conditions

$$M_{-1}[c_0, c_1; g_i] = 2, \quad M_0[c_0, c_1; g_i] = 0, \qquad (3.32)$$

These two equations follow by contraction of the contour in the last integral in (3.29) to $C$.

### 3.3 Complete solution $W_0(z_1, ..., z_s)$

In principle the solution at spherical level is given by (3.31)-(3.32). These equations define $W_0(z)$ and we can apply the loop inserting operator to obtain any multi-loop correlator. Quite surprisingly one can obtain an explicit formula if we change the matrix model slightly. Instead of Hermitian matrices we use general complex matrices and consider a potential:

$$V(\phi^\dagger \phi) = \sum_{n=1}^{\infty} \frac{g_n}{n} \text{Tr}\,(\phi^\dagger \phi)^n. \qquad (3.33)$$

The one-loop correlator is given by:

$$W_0(z) \equiv \frac{1}{N} \sum_n \frac{\langle \text{Tr}\,(\phi^\dagger \phi)^n \rangle}{z^{2k+1}}. \tag{3.34}$$

This model has a surface representation [16]. $\text{Tr}\,(\phi^\dagger \phi)^n$ represents a $2n$-gon where the boundary links have alternating black and white colors, corresponding the $\phi$ and $\phi^\dagger$. The effect of Gaussian integration with respect to the complex matrices is to glue together such "checker-board" polygons just as Hermitian matrices glued together ordinary polygons. The only additional rule is that only "white" and "black" links can be glued together since $\langle \phi^\dagger \phi \rangle \neq 0$ while $\langle \phi^2 \rangle = 0$ and $\langle \phi^\dagger \phi^\dagger \rangle = 0$. Such short distance differences in gluing should be unimportant in the continuum limit.

We can write down the loop equations for this model [12]. They are similar to the ones already considered, except that we now have a symmetry with respect to $\phi \to -\phi$, i.e. $c_0 = -c_1$ and the formulas above simplify:

$$W_0(z) = \frac{1}{2}\left(V'(z) - M(z)\sqrt{z^2 - c_1^2}\right), \quad M(z) = \oint_{C_\infty} \frac{d\omega}{4\pi i} \frac{\omega V'(\omega)}{(\omega^2 - z^2)\sqrt{\omega^2 - c_1^2}} \tag{3.35}$$

$$M(z) = \sum_{k=1}^\infty M_k[c_1, g_i](z^2 - c_1^2)^{k-1}, \quad M_k[c_1, g_i] = \oint_C \frac{d\omega}{4\pi i} \frac{\omega V'(\omega)}{(\omega^2 - c_1^2)^{k+1/2}}, \tag{3.36}$$

and the position $c_1$ of the cut is determined by

$$M_0[c_1, g_i] = 2. \tag{3.37}$$

*While these formulas look rather complicated it is a pleasant surprise that things simplify a lot for the higher loop correlators.* In order to apply the loop insertion operator we note that

$$\frac{d}{dV(z)} \equiv -\sum_k \frac{k}{z^{2k+1}} \frac{d}{dg_k} = \frac{\partial}{\partial V(z)} - \frac{c_1^2}{M_1(z^2 - c_1^2)^{3/2}} \frac{d}{dc_1^2}. \tag{3.38}$$

where

$$\frac{\partial}{\partial V(z)} \equiv -\sum_k \frac{k}{z^{2k+1}} \frac{\partial}{\partial g_k}, \quad \frac{\partial V'(\omega)}{\partial V(z)} = \frac{2\omega z}{z^2 - \omega^2}. \tag{3.39}$$

It is now straightforward, though tedious, to apply the loop insertion operator. We find [12]:

$$W_0(z_1, z_2) = \frac{1}{4(z_1^2 - z_2^2)^2}\left[z_2^2 \sqrt{\frac{z_1^2 - c_1^2}{z_2^2 - c_1^2}} + z_1^2 \sqrt{\frac{z_2^2 - c_1^2}{z_1^2 - c_1^2}} - 2z_1 z_2\right]. \tag{3.40}$$

$$W_0(z_1, z_2, z_3) = \frac{c_1^4}{16 M_1} \frac{1}{\sqrt{(z_1^2 - c_1^2)(z_2^2 - c_1^2)(z_3^2 - c_1^2)}}. \tag{3.41}$$

$$W_0(z_1, \ldots, z_s) = \left(\frac{1}{M_1} \frac{d}{dc_1^2}\right)^{s-3} \frac{1}{2c_1^2 M_1} \prod_{k=1}^s \frac{c_1^2}{2(z_k^2 - c_1^2)^{3/2}}. \tag{3.42}$$

Note that the above formulas are valid for *any* potential $V$. All dependence on the coupling constants is hidden in $M_1$ and $c_1$.

## 3.4 The scaling limit

In the case of the simplest potential ($\operatorname{Tr}\phi^3$ for the Hermitian matrix model, $\operatorname{Tr}(\phi^\dagger\phi)^2$ for the complex matrix model), we have one independent coupling constant $g$ if we fix the coupling constant in front of the Gaussian term. Eq. (3.9) gives the relation between the bare cosmological coupling constant $\mu$ and the coupling constant $g$ of the matrix models. We have seen that there is a critical $\mu_c$ such that the continuum limit should be taken for $\mu \to \mu_c$. Corresponding to $\mu_c$ there will be a $g_c$. If we introduce the lattice spacing $a$, the area of our universe for a given triangulation is given by $N_T a^2$, $N_T$ being the number of triangles in $T$ and as already discussed it is natural to introduce the *renormalized* cosmological constant $\Lambda$ by

$$\mu - \mu_c = \Lambda a^2, \quad \text{i.e.} \quad g_c - g \sim g_c \Lambda a^2. \tag{3.43}$$

Let us for simplicity consider the complex matrix model. The density of eigenvalues $\rho(\lambda)$ is given by (3.27) and (3.36):

$$\rho(\lambda) \sim M(\lambda)\sqrt{c_1^2 - \lambda^2}, \quad M(\lambda) = M_1 + M_2(c_1^2 - \lambda^2). \tag{3.44}$$

The only non-analytic behavior is associated with the endpoints of the distribution where $|\lambda| \to c_1$. If $M_1[g]$ is positive the behavior will be identical to that of the Gaussian model. We conclude that the critical point $g_c$ is determined by:

$$M_1[c_1(g_c), g_c] = 0. \tag{3.45}$$

A glance on $W_0(z_1, ..., z_s)$ corroborates this observation since it will be singular precisely when $M_1 = 0$. Let us move slightly away from the critical point by scaling $g$ according to (3.43):

$$g = g_c(1 - \Lambda a^2) \tag{3.46}$$

Corresponding to this change there will be a change $c_1^{(c)2} \to c_1^{(c)2} - \delta c_1^2$. It can be calculated directly from (3.37) by expanding $M_0[c_1, g]$ and we get:

$$(\delta c_1^2)^2 = \frac{16}{3M_2(g_c)}\Lambda a^2 + \mathcal{O}(a^3) \quad \text{i.e.} \quad c_1^2 = c_1^{(c)2} - a\sqrt{\Lambda} + \mathcal{O}(a^2), \tag{3.47}$$

where we have absorbed a constant $k^2 = 16/3M_2(g_c)$ in the definition of the cosmological constant $\Lambda$. In addition $M_1[c_1, g_i]$ will now be different from zero:

$$M_1[c_1, g_i] = \frac{3M_2(g_c)}{2}\delta c_1^2 + O(a^2). \tag{3.48}$$

From the explicit formulas for the multi-loop correlators it follows that the complex variable $z$ appears in the combination $z^2 - x^2$ and it is natural to introduce a scaling of $z^2$:

$$z^2 = z_c^2 + a Z, \quad z_c = c_1^{(c)} \tag{3.49}$$

With this notation we can write in the scaling limit :

$$M(z) = M_1 + M_2(z^2 - x^2) = aM_2(g_c)(Z - \frac{1}{2}\sqrt{\Lambda}) + O(a^2) \tag{3.50}$$

$Z$ serves the same role in the scaling limit as $z$ at the discretized level: knowing $W_0(z_1,...z_s)$ allows us to reconstruct the multi-loop correlators consisting of discretized boundary loops of length $n_1,...,n_s$ by multiple contour integration. In the scaling limit the physical length of these loops will be $L_i = n_i a$, i.e. they will go to zero. If we want *genuine macroscopic loops* we have to scale $n_i$ to $\infty$ at the same time as $a \to 0$ such that $L_i$ is constant. By substituting (3.49) and (3.50) in the expressions for $W_0(z_1,...z_s)$ we get an expression $W_0(Z_1,...,Z_s; \Lambda)$ and we can reconstruct the corresponding multi-loop amplitude $W_0(L_1,....,L_s; \Lambda)$ by an inverse Laplace transform. The contour integration is changed into an inverse Laplace transform since the cut $[-c_0, c_1]$ (or $[c_0, c_1]$ for the Hermitian matrix model) by the substitution (3.49) and (3.50) is changed into a cut $[-\infty, \sqrt{\Lambda}]$. The contour integration around the cut can now be deformed:

$$\oint_{C_i} \prod_{i=1}^{s} dz_i\, z^{2n_i}\, W_0(z_1, \ldots, z_s) \to \tag{3.51}$$

$$z_c^{2(n_1 + \cdots + n_s)} \int_{c-i\infty}^{c+i\infty} \prod_{i=1}^{s} dZ_i\, e^{L_i Z_i}\, W_0(Z_1, \ldots, Z_s; \Lambda), \quad c > -\sqrt{\Lambda},$$
$$\tag{3.52}$$

since we have

$$z^{2n} = z_c^{2n}(1 + aZ)^{L/a} \sim z_c^{2n} e^{ZL}, \quad L \equiv n/a. \tag{3.53}$$

The highly divergent factor $z_c^{2(L_1 + \cdots + L_s)/a}$ is a wave function renormalization of the macroscopic boundaries. It is to be expected since it is possible in the continuum to add a term

$$S_{\mathcal{M}} = \lambda \int_{\partial \mathcal{M}} dl = \lambda \int d\xi\, g^{\frac{1}{4}}(\xi) \tag{3.54}$$

to the action. This is just the induced one-dimensional gravity on the boundary of the manifold. Since $\lambda$ has the dimension of mass we expect in the discretized version that the bare coupling constant will undergo an additive renormalization (like the cosmological constant itself), i.e. we will have to cancel a term like $z_c^{2n}$.

From eq. (3.42) we immediately get the generating functional for macroscopic multi-loop amplitudes [12]:

$$W_0(z_1, \ldots, z_s) \sim a^{5 - 7s/2} W(Z_1, \ldots, Z_s; \Lambda), \tag{3.55}$$

$$W(Z_1, \ldots, Z_s; \Lambda) = \frac{d^{s-3}}{d\Lambda^{s-3}} \prod_{k=1}^{s} \frac{1}{(Z_k + \sqrt{\Lambda})^{3/2}}, \quad s \geq 3. \tag{3.56}$$

The expressions for $W(Z, \Lambda)$ and $Z(Z_1, Z_2; \Lambda)$ are slightly more complicated since $W_0(z)$ contains a non-universal part $(V'(z)/2)$. and we will not give them here.

We can calculate the inverse Laplace transform (3.51)

$$W(L_1, \ldots, L_s; \Lambda) =$$
$$\int_{c-i\infty}^{c+i\infty} \prod_{i=1}^{s} dZ_i\, e^{Z_i L_i}\, W(Z_1, \ldots, Z_s; \Lambda) = \frac{d^{s-3}}{d\Lambda^{s-3}} \frac{e^{-\sqrt{\Lambda}(L_1 + \cdots + L_s)}}{\sqrt{L_1 \cdots L_s}}.$$
$$\tag{3.57}$$

*These are the Hartle-Hawkings wave functionals (2.3) of two-dimensional quantum gravity.*

# 4 The two-point function

Let us now turn to the problem of calculating the two-point function of two-dimensional quantum gravity as a function of the geodesic distance.

## 4.1 Formulation of the combinatorial problem

Let $F(x;g)$ denote the generating functional for the number $\mathcal{N}_{n,l}$ of triangulations of the disk with $n$ triangles and one boundary of length $l$. It is given by $F(x;g) = W_0(1/x;g)/x$, where $W_0(z;g)$ is the one-loop function for a cubic matrix model calculated in the last section[5]:

$$F(x;g) = \sum_{n,l} \mathcal{N}_{n,l} g^n x^l = \sum_{l=0}^{\infty} x^l F(l;g)$$
$$= \frac{1}{2}\left(\frac{1}{x^2} - \frac{g}{x^3}\right) + f(x;g), \quad (4.1)$$

$$f(x;g) = \frac{g}{2x}\left(\frac{1}{x} - c_2\right)\sqrt{\left(\frac{1}{x} - c_1\right)\left(\frac{1}{x} - c_0\right)}, \quad (4.2)$$

where $c_i$ are functions of $g$ determined by the requirement[6] $F(x;g) \sim \mathcal{O}(1)$ for $x \to 0$. We have $F(l;g) = \langle \mathrm{Tr}\,\phi^l/N \rangle$ in the matrix model notation of the last section and it has the interpretation as the generating functional for all triangulations with a boundary consisting of $l$ links of which one is marked. For obvious reasons we denote $F(x;g)$ the disk amplitude. As usual we have $g = e^{-\mu}$ where $\mu$ is the cosmological constant. Corresponding to the critical value $\mu_c$ we have a critical value $g_c$ as explained in the last section and for $\mu$ close to $\mu_c$ we have

$$\triangle \mu \equiv \mu - \mu_c = \triangle g/g_c. \quad (4.3)$$

One has $c_0 < 0 < c_1 < c_2$ as long as $g < g_c$. The only thing we need to know here is that for $g = g_c$ we have $c_1 = c_2$ (which we denote $1/x_c$, while for $g$ close to $g_c$, i.e. $\mu$ close to $\mu_c$ we have (as follows directly from (3.47)-(3.50))

$$c_2(\mu) = 1/x_c + \frac{\alpha}{2}\sqrt{\triangle \mu} + \mathcal{O}(\triangle \mu) \quad (4.4)$$
$$c_1(\mu) = 1/x_c - \alpha\sqrt{\triangle \mu} + \mathcal{O}(\triangle \mu) \quad (4.5)$$
$$c_0(\mu) = c_0(\mu_c) + \mathcal{O}(\triangle \mu) \quad (4.6)$$

where $\alpha = 4/3^{\frac{1}{4}}$.

Let us consider triangulations with two boundary loops[7] $l_1$ and $l_2$ separated by a geodesic distance $r$ (to be defined below). We call $l_1$ the entrance loop and $l_2$ the exit loop and it is convenient to view one of the links of $l_1$ as being marked. For a given triangulation we have already defined the geodesic distance $d(l, l_1)$ between a link $l$ and a set of links $l_1$ to be $\min_{l' \in l_1} d(l, l')$. We say that a loop $l_2$ has a geodesic distance $r$ to another loop $l_1$ if *all* links $l \in l_2$ have a geodesic distance $r$ to $l_1$. Note that the definition is not symmetric in $l_1$ and $l_2$. However, it is the natural definition for the application we have in mind, as will be clear from the composition law (4.8). We denote the class of such triangulations $\mathcal{T}(r, l_1, l_2)$. The geometry of a triangulation in $\mathcal{T}(r, l_1, l_2)$ will be that of a cylinder with "height" $r$ (see fig. 6). $\mathcal{T}(r, l_1, l_2)$ is related to the class of triangulations $\mathcal{T}(l_1)$ with a boundary of length $l_1$ and one marked link as shown in fig. 7. As can be seen from fig. 7 the elements in $\mathcal{T}(r, l_1, l_2)$ are obtained by chopping off a disk $T'$ from the disk $T \in \mathcal{T}(l_1)$ such that a loop of length $l_2$ and geodesic distance $r$ to $l_1$ is created. As illustrated in the figure the loop $l_2$ might only be one of many loops in $T$ which have a geodesic distance $r$ to $l_1$.

---

[5] We find it convenient in this section to change from the variable $z$ used in the last section to $x = 1/z$.
[6] The constants $c_i$ are determined by eqs. (3.29) and (3.32) of the last section.
[7] We use the same notation $l_1$ and $l_2$ for the set of boundary links and the number of boundary links.

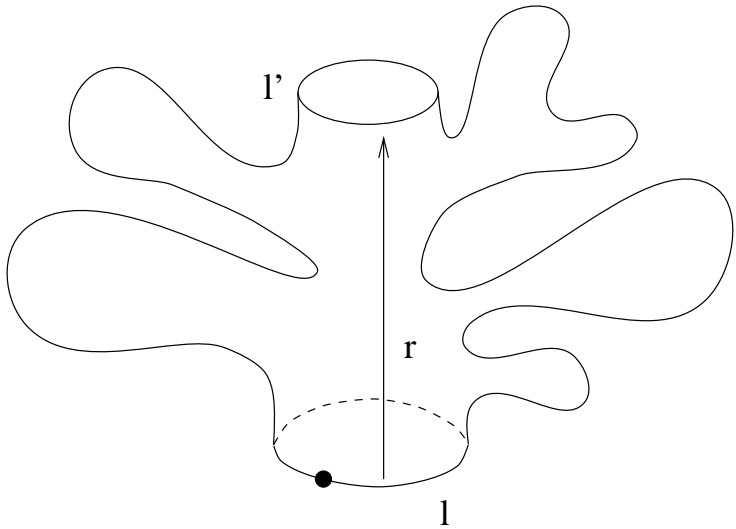

Figure 6: A typical graph contributing to $G_\mu(l, l'; r)$. A blob represents a marked link.

We can now define a generalization of the two-point function $G_\mu(r)$ to the two-loop function, where the loops are separated by a geodesic distance $r$:

$$G_\mu(l_1, l_2; r) = \sum_{T \in \mathcal{T}(r, l_1, l_2)} e^{-\mu N_T}, \qquad (4.7)$$

Like the two-point function, we expect from general considerations that the two-loop function falls of exponentially for $r \to \infty$ with the same exponent as the two-point function. In addition one has the following fundamental composition law which comes from the fact that a cylinder of height $r_1 + r_2$ has a unique decomposition into cylinders of height $r_1$ and $r_2$, respectively (see fig. 8):

$$G_\mu(l_1, l_2; r_1 + r_2) = \sum_{l=1}^{\infty} G_\mu(l_1, l; r_1) \, G_\mu(l, l_2; r_2). \qquad (4.8)$$

This law expresses nothing but the summation over intermediate states since these can depend only on the equivalence classes of the one-dimensional (induced) metrics, and the equivalence classes of one-dimensional metrics are uniquely characterized by the length of the loop. Using (4.8) we can in principle find $G_\mu(l_1, l_2; r)$ if we know $G_\mu(l, l'; 1)$ and it is seen that $G_\mu(l, l'; 1)$ via (4.8) acts as a *transfer matrix*. *It is possible to find the transfer*

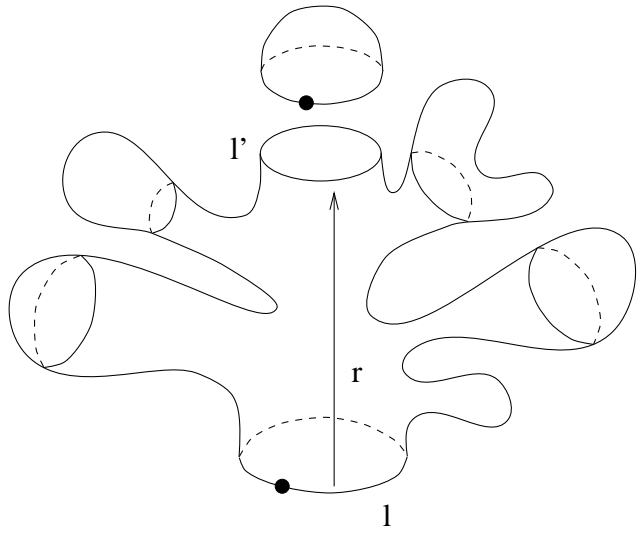

Figure 7: Illustration of how to obtain a graph in $\mathcal{T}(r,l,l')$ from a graph in $\mathcal{T}(l)$ by chopping off a disk with the boundary length $l'$. As shown there can be many loops of length $l'$ and geodesic distance $r$ from $l$ in a specific graph belonging to $\mathcal{T}(l)$.

*matrix $G_\mu(l,l';1)$ by purely combinatorial arguments.* As for the disk amplitude and the multiloop correlators it is convenient to introduce the generating functional:

$$G_\mu(x,y;r) = \sum_{l_1,l_2=1}^{\infty} x^{l_1} y^{l_2} G_\mu(l_1,l_2;r), \quad G_\mu(x,y;r=0) = \frac{xy}{1-xy}. \tag{4.9}$$

The last equation follows from the observation that

$$G_\mu(l_1,l_2;r=0) = \delta_{l_1,l_2}. \tag{4.10}$$

We note that $G_\mu(l_1,l_2;r)$ can be reconstructed from the knowledge of $G_\mu(x,y;r)$:

$$G_\mu(l_1,l_2;r) = \oint_{C_x} \frac{dx}{2\pi i x} x^{-l_1} \oint_{C_y} \frac{dy}{2\pi i y} y^{-l_2} G_\mu(x,y;r) \tag{4.11}$$

where the contours $C_x$ and $C_y$ surround the origin and avoid the cuts of $G_\mu(x,y;r)$. Further it should be noted that the fundamental composition law (4.8) can be rewritten as:

$$G_\mu(x,y;r_1+r_2) = \oint_C \frac{dz}{2\pi i z} G_\mu(x,\frac{1}{z};r_1) G_\mu(z,y;r_2). \tag{4.12}$$

This allows us to write:

$$G_\mu(x,y;r+1) = \oint_C \frac{dz}{2\pi i z} G_\mu(x,\frac{1}{z};1) G_\mu(z,y;r). \tag{4.13}$$

Our main problem will be to find an explicit expression for $G_\mu(x,y;1)$. This problem was first solved in [17, 18].

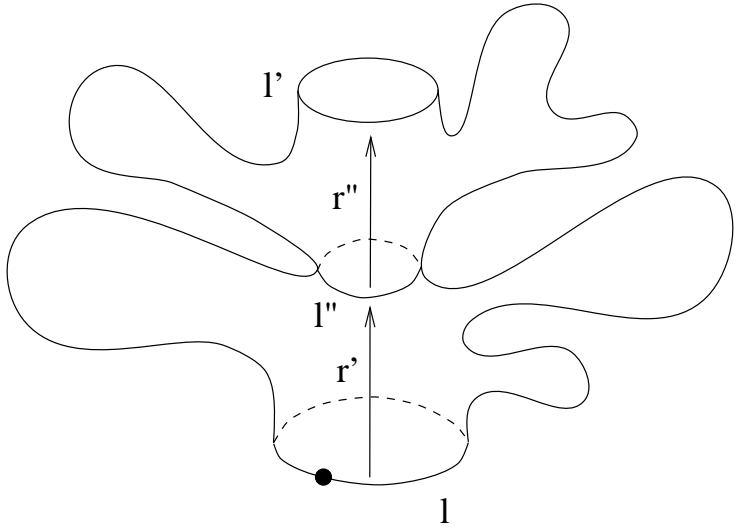

Figure 8: A typical graph which explains the composition law of $G_\mu(l, l'; r)$.

## 4.2 Solution of the combinatorial problem

Let $G_\mu^{(0)}(x, y)$ denote the transfer matrix where the entrance loop is not marked. In the counting of different configurations the marking of the entrance loop will convert one unmarked loop into $l_1$ marked loops, i.e. $G_\mu^{(0)}(x, y)$ is related $G_\mu(x, y; 1)$ by:

$$G_\mu(x, y; 1) = x \frac{\partial}{\partial x} G_\mu^{(0)}(x, y). \tag{4.14}$$

The transfer matrix $G_\mu^{(0)}(x, y)$ is the sum of all possible graphs connecting loops separated by a geodesic distance $r = 1$. For each triangle there will be a factor $g = e^{-\mu}$. In fig. 9 we show a typical graph. One should notice that the graphs of this kind are obtained by combining around the boundary four types of graphs, △ , ▽ , △ , ▯ (where blobs ⊘ denote disk topology). In this way $G_\mu^{(0)}(x, y)$ can be written as

$$\begin{aligned} G_\mu^{(0)}(x, y) &= \sum_{n=1}^{\infty} \frac{1}{n} (\,\triangle + \triangledown + \triangle + \blacksquare\,)^n \\ &= -\log(1 - \triangle - \triangledown - \triangle - \blacksquare\,), \end{aligned} \tag{4.15}$$

where the factor $1/n$ comes from the cyclic symmetry. The explicit expressions for △ , ▽ , △ , ▯ are as follows:

$$\triangle = gxy^2, \tag{4.16}$$

$$\triangledown = gx^2 F(x; g) y, \tag{4.17}$$

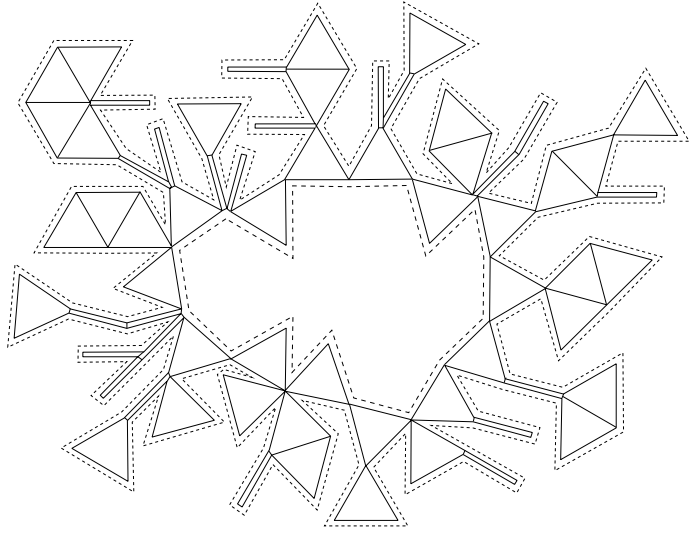

Figure 9: A typical graph contributing to $G_\mu^{(0)}(x,y)$. a thin broken line represents an entrance loop and a thick broken line represents an exit loop.

$$\triangle = gx^2 \frac{F(x;g) - 1}{x}, \qquad (4.18)$$

$$\bigcirc = x^2 F(x;g), \qquad (4.19)$$

since $\triangle$ contains two links from the exit loop, one from the entrance loop and one triangle, i.e. a factor $y^2xg$ and $\bigtriangledown$ contains one link from the exit loop, two links plus any disk graph with one marked vertex (where the two links join the disk amplitude) and one additional triangle, i.e. a factor $yx^2 F(x;g)g$. Similar explanations are immediately obtained for $\triangle$ and $\bigcirc$. Substituting this into (4.15), we get (with $g \equiv e^{-\mu}$)

$$G_\mu^{(0)}(x,y) = -\log\left(1 - gxy^2 - gx^2 F(x;g)y - gx(F(x;g) - 1) - x^2 F(x;g)\right). \qquad (4.20)$$

The continuum limit of the above expression is taken in the same way as when we discussed the continuum limit of multiloop correlators $W_0(z_1, \ldots, z_s)$ in the last section, the variable $x$ of the entrance loop being analogous to the variable $1/z$ of the loop correlators, as is clear from the relation between $F(x;g)$ and $W_0(z)$. On the other hand it follows from the composition law (4.12) that the continuum limit of $x$ and $y$ should be taken around values which are inverse to each other, i.e. the variable $y$ of the exit loop is similar to $z$ itself. Consequently we perform the expansion:

$$x = x_c(1 - aX), \qquad y = \frac{1}{x_c}(1 - aY), \qquad g = g_c(1 - \Lambda a^2), \qquad (4.21)$$

where $x_c = 1/z_c$. In addition we know from the properties of the two-point function that the geodesic distance at the continuous level, $R$, should be related to the geodesic distance

$r$ at the discrete level as in (2.29), i.e. introducing the lattice scaling $a(\mu)$:

$$R \sim r\, a^{2\nu}, \tag{4.22}$$

and that the continuum two-point function is related to the discretized two-point function by a multiplicative renormalization (2.32). Let us assume that we also have a multiplicative renormalization of the two-loop function $G_\mu(x,y;r)$:

$$G_\Lambda(X,Y;R) = a^{\delta_G} G_\mu(x,y;r), \tag{4.23}$$

where $\delta_G$ is a real number. As for the multi-loop correlators $W_0(z_1,...,z_s)$ the contour integrals in the composition law (4.12) go into inverse Laplace transforms by the substitution (4.21) and we get:

$$G_\Lambda(X,Y;R_1+R_2) = a^{1-\delta_G} \int_{-i\infty}^{i\infty} \frac{dZ}{2\pi i} G_\Lambda(X,-Z;R_1) G_\Lambda(Z,Y;R_2), \tag{4.24}$$

i.e. we find $\delta_G = 1$. Substituting (4.21) into eq. (4.20), we obtain

$$G_\mu^{(0)}(x,y) = -\log\Big[\text{const.}\,\{(X+Y)a - \mathcal{F}_\Lambda(X)a^{3/2} + \mathcal{O}(a^2)\}\Big], \tag{4.25}$$

where we introduced the following notation for the "scaling" part $f(x;g)$ of the disk amplitude given by eq. (4.2):

$$f_\mu(x) \equiv f(x;g) = \mathcal{F}_\Lambda(X)a^{3/2} + \mathcal{O}(a^2) \tag{4.26}$$

$$\mathcal{F}_\Lambda(X) = \tilde{\alpha}(X - \tfrac{1}{2}\sqrt{\Lambda})\sqrt{X + \sqrt{\Lambda}}, \tag{4.27}$$

and where $\tilde{\alpha} = \sqrt{2/(9\sqrt{3}-1)}$. Thus, we obtain the transfer matrix at the continuous level,

$$G_\Lambda(X,Y;R=a^{2\nu}) = aG_\mu(x,y;r=1) = \frac{1}{X+Y} - a^{1/2}\frac{\partial}{\partial X}\left[\frac{\mathcal{F}_\Lambda(X)}{X+Y}\right] + \mathcal{O}(a). \tag{4.28}$$

Note that the leading term in (4.28) is the Laplace transformed delta function $\delta(L-L')$. By substituting (4.28) into (4.24) for $r_1 = 1$ and $r_2 = r$, we obtain

$$\begin{aligned}
G_\Lambda(X,Y;R+a^{2\nu}) &= \int_{-i\infty}^{i\infty} \frac{dZ}{2\pi i}\Big(\frac{1}{X-Z} - a^{1/2}\frac{\partial}{\partial X}\frac{\mathcal{F}_\Lambda(X)}{X-Z} + \mathcal{O}(a)\Big)G_\Lambda(Z,Y;R) \\
&= G_\Lambda(X,Y;R) - a^{1/2}\frac{\partial}{\partial X}[\mathcal{F}_\Lambda(X)G_\Lambda(X,Y;R)] + \mathcal{O}(a). \quad (4.29)
\end{aligned}$$

*We only get a non-trivial continuum limit if $2\nu = 1/2$, i.e. the Hausdorff dimension $d_h = 1/\nu = 4$.*

If we use $\nu = 1/4$ we obtain the continuum differential equation [17]:

$$\frac{\partial}{\partial R}G_\Lambda(X,Y;R) = -\frac{\partial}{\partial X}[\mathcal{F}_\Lambda(X)G_\Lambda(X,Y;R)]. \tag{4.30}$$

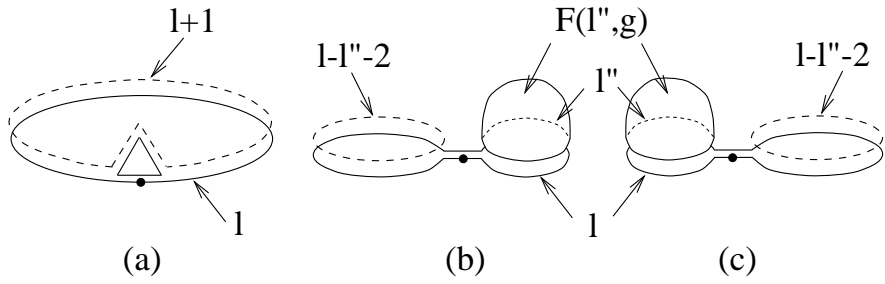

Figure 10: Fundamental peeling decompositions. (a) removes a triangle while (b) and (c) remove a two-folded part. The solid line with the mark represents the entrance loop and the dashed line the exit loop after one peeling step.

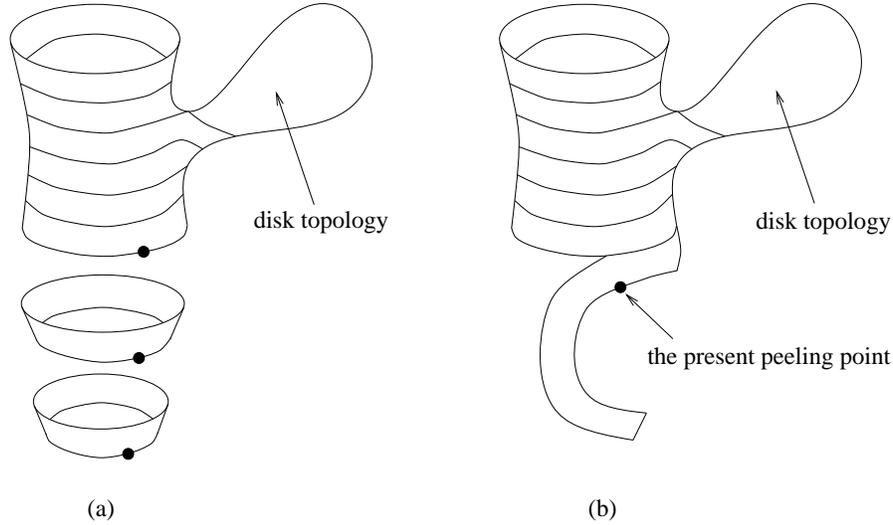

Figure 11: Decomposition of the surface by (a) slicing and (b) peeling.

Let us check the universality by deriving the differential equation (4.30) differently. At the discretized level the procedure outlined above can be viewed as a "slicing decomposition" of the triangulations. Starting from the entrance loop we "slice" the triangulations in cylinders of thickness $r = 1$. Let us instead consider a different set of "deformations" which can bring us from the entrance loop to the exit loop, the so called "peeling decomposition" [18]. In this decomposition we start at the marked link at the entrance loop and remove the triangle containing this link. We might have a situation where the marked link is not associated with a triangle, but is folded with another link. In this case we simply remove both links. These three fundamental steps in the peeling decomposition are shown in fig.10. Successive application of these steps is like peeling an apple (see fig.11), in contrast to the situation above which was like slicing the apple and the individual steps in the peeling decomposition can be viewed as $1/l$ of the slicing decomposition, where $l$ is the length of the boundary.

The fundamental equation satisfied by $G_\mu(l, l'; r)$ by a one-step peeling will be:

$$G_\mu(l, l'; r) \to g G_\mu(l+1, l'; r) + 2 \sum_{l''=0}^{l-2} F(l''; g) G_\mu(l - l'' - 2, l'; r). \tag{4.31}$$

If we identify the deformation (4.31) with $(1/l)$-step of the slicing decomposition, we find

$$\frac{1}{l} \frac{\partial}{\partial r} G_\mu(l, l'; r) = g G_\mu(l+1, l'; r) + 2 \sum_{l''=0}^{l-2} F_\mu(l'') G_\mu(l - l'' - 2, l'; r) - G_\mu(l, l'; r). \tag{4.32}$$

Taking the discrete Laplace transform of (4.32) we get:

$$\frac{\partial}{\partial r} G_\mu(x, y; r) = x \frac{\partial}{\partial x} \left( 2x^2 f_\mu(x) G_\mu(x, y; r) \right), \quad G_\mu(x, y; r=0) = \frac{xy}{1 - xy}. \tag{4.33}$$

In this case it is manifest that one can obtain (4.30) after taking the continuum limit [18].

The peeling decomposition has a number of advantages compared to the slicing decomposition. One can easily use it to construct a string field theory [18] and from the point of view of this article it has the advantage that we have a simple differential equation even at the level of discretized loop length. This allows us to avoid certain ambiguities of first taking the discretized loop length to infinity by the substitution (4.21) and next taking it to zero in order to construct the two-point function.

## 4.3 Solution of the differential equation

We now discuss the solution [8] to eqs. (4.33). It is readily found since we have a first order partial differential equation:

$$G_\mu(x, y; r) = \frac{\hat{x}^2 f_\mu(\hat{x})}{x^2 f_\mu(x)} \frac{\hat{x} y}{1 - \hat{x} y}. \tag{4.34}$$

Here $\hat{x}(x, r)$ is the solution to the characteristic equation of the partial differential equation (4.33). The integral of the characteristic equation is

$$r = \int_x^{\hat{x}(x,r)} \frac{dx'}{2 x'^3 f_\mu(x')} = \left[ \frac{1}{\delta_0} \sinh^{-1} \sqrt{\frac{\delta_1}{1 - c_2 x'} - \delta_2} \right]_{x'=x}^{x'=\hat{x}(x,r)}, \tag{4.35}$$

and this expression can be by inverted to give:

$$\hat{x}(x, r) = \frac{1}{c_2} - \frac{\delta_1}{c_2} \frac{1}{\sinh^2 \left( \delta_0 r + \sinh^{-1} \sqrt{\frac{\delta_1}{1 - c_2 x} - \delta_2} \right) + \delta_2}, \tag{4.36}$$

where $\delta_0$, $\delta_1$ and $\delta_2$ are all positive and defined by

$$\delta_0 = \frac{g}{2} \sqrt{(c_2 - c_1)(c_2 - c_0)} = \mathcal{O}((\triangle\mu)^{\frac{1}{4}}), \tag{4.37}$$

$$\delta_1 = \frac{(c_2 - c_1)(c_2 - c_0)}{c_2 (c_1 - c_0)} = \mathcal{O}(\sqrt{\triangle\mu}), \tag{4.38}$$

$$\delta_2 = -\frac{c_0 (c_2 - c_1)}{c_2 (c_1 - c_0)} = \mathcal{O}(\sqrt{\triangle\mu}). \tag{4.39}$$

It is readily checked that $\hat{x} \to 1/c_2$ for $r \to \infty$ and $\hat{x}(x, r=0) = x$.

In principle we can calculate $G_\mu(l_1, l_2; r)$ from eqs. (4.12), (4.34) and (4.36). Here let us only verify that the exponential decay of $G_\mu(l_1, l_2; r)$ is independent of $l_1$ and $l_2$. For $r \to \infty$ one gets

$$G_\mu(l_1, l_2; r) = \text{const.} \delta_0 \delta_1 \, e^{-2\delta_0 r} + \mathcal{O}(e^{-4\delta_0 r}). \tag{4.40}$$

where *const.* is a function of $\mathcal{O}(1)$ which depends on $c_0, c_1, c_2, l_1$ and $l_2$.

We can express the two-point function $G_\mu(r)$ in terms of the two-loop function and the one-loop function. Let us consider a marked link. For a given triangulation we can systematically work our way out to the links having a distance $r$ from the marked link by peeling off layers of triangles having the distances $1, 2, \ldots, r$ to the marked link. After $r$ steps we have a boundary consisting of a number of disconnected boundary loops, all with a distance $r$ to the marked link. One of these is the exit loop described by the two-loop function and we get the two-point function by closing the exit loop of length $l_2$ by multiplying the two-loop function $G_\mu(l_1 = 1, l_2; r)$ by the one-loop function[8] $F_\mu(l_2)$ and performing the sum over $l_2$, i.e., as shown in fig. 12,

$$\begin{aligned} G_\mu(r) &= \sum_{l_2=1}^{\infty} G_\mu(l_1 = 1, l_2; r) \, l_2 F_\mu(l_2) \\ &= \frac{\partial}{\partial x} \oint_{C_y} \frac{dy}{2\pi i y} G_\mu\left(x, \frac{1}{y}; r\right) y \frac{\partial}{\partial y} F_\mu(y) \bigg|_{x=0} \\ &= \frac{\partial}{\partial x} F_\mu(\hat{x}) \bigg|_{x=0} = \frac{1}{g} \frac{\partial}{\partial r} F_\mu(\hat{x}) \bigg|_{x=0}. \end{aligned} \tag{4.41}$$

As long as $c_2 - c_1$ is small and $r$ is larger than a few lattice spacings, we get[8]:

$$G_\mu(r) = \text{const.} \delta_0 \delta_1 \frac{\cosh(\delta_0 r)}{\sinh^3(\delta_0 r)} \left(1 + \mathcal{O}(\delta_0)\right). \tag{4.42}$$

Formula (4.42) shows how to take the scaling limit: Let us return to the original formulation and write in the limit $\triangle\mu \to 0$:

$$G_\mu(r) = \text{const.} \, (\triangle\mu)^{3/4} \frac{\cosh\left[(\triangle\mu)^{\frac{1}{4}} \beta r\right]}{\sinh^3\left[(\triangle\mu)^{\frac{1}{4}} \beta r\right]}. \tag{4.43}$$

where *const.* and $\beta$ are positive constants of $\mathcal{O}(1)$ ($\beta = \sqrt{6} g_c$).

We conclude the following:

1. $G_\mu(r)$ falls of like $e^{-2(\triangle\mu)^{\frac{1}{4}} \beta r}$ for $r \to \infty$, i.e. the critical exponent $\nu = \frac{1}{4}$ and the Hausdorff dimension $d_H = 4$.

2. $G_\mu(r)$ behaves like $r^{-3}$ for $1 \ll r \ll \triangle\mu^{-\frac{1}{4}}$, i.e. the scaling exponent $\eta = 4$.

3. The pre-exponential factor to $e^{-2(\triangle\mu)^{\frac{1}{4}} \beta r}$ is $(\triangle\mu)^{\eta-1}$, i.e. precisely the factor needed in order to take a continuum limit according to the general discussion following eq. (2.27).

---

[8]To be more precise we have to multiply the two-loop function $G_\mu(l_1, l_2; r)$ by $l_2$ since the exit loop is unmarked and we can glue the marked one-loop cap to it in $l_2$ ways.

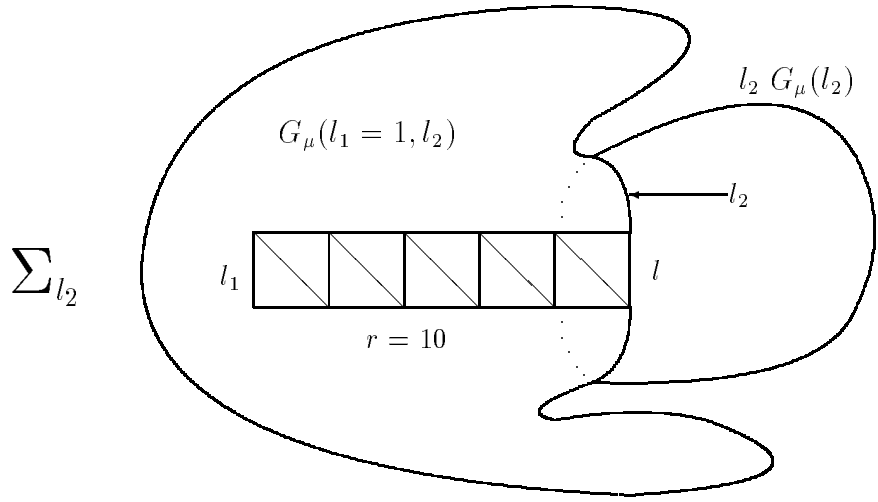

Figure 12: The 2-point function represented as a summation over 2-loop functions times 1-loop functions.

4. From Fisher's scaling relation we get $\gamma = \nu(2-\eta) = -1/2$. This well known result can of course also be derived directly from

$$\chi(\mu) = \sum_{r=1}^{\infty} G_\mu(r) = \text{const.} - c^2(\triangle\mu)^{\frac{1}{2}} + \cdots \qquad (4.44)$$

by use of (4.43), but it should be clear that the explicit calculation in (4.44) is nothing but a specific example of the general calculation used in proving Fisher's scaling relation. What is somewhat unusual compared to ordinary statistical systems is that the anomalous scaling dimension $\eta > 2$. $\eta = 0$ is the ordinary free field result, while $\eta = 2$ is the infinite temperature limit, and for statistical systems we expect $\eta < 2$.

5. Any two-loop function $G_\mu(l_1, l_2; r)$ has the same behavior as $G_\mu(r)$ as long as $l_1, l_2$ stay finite as $\triangle\mu \to 0$.

It is clear that we could have taken the continuum limit almost at any point in the above calculations and we get

$$G(R; \Lambda) = \lim_{a \to 0} (\sqrt{a})^{\eta-1} G_\mu(r) \sim (\Lambda)^{3/4} \frac{\cosh[(\Lambda)^{\frac{1}{4}} R]}{\sinh^3[(\Lambda)^{\frac{1}{4}} R]}. \qquad (4.45)$$

where we have rescaled $R$ with a factor $\beta$ compared to the definition $R = r\, a^{2\nu}$ in eq. (4.22). The factor in front of $G_\mu(r)$ is the usual "wave function renormalization" present in the path integral representation of the propagator and the unusual power $M = 2\Lambda^{1/4}$ of the "mass" is due to $d_H = 4$.

# 5 Euclidean quantum gravity in $d > 2$

## 5.1 Basic questions in Euclidean quantum gravity

The moment we address Euclidean quantum gravity in dimensions higher than two a frightening number of basic questions appears. Let us just list some of them:

(1): How do we cure the unboundedness of the Einstein-Hilbert action in $d > 2$?

(2): Does the non-renormalizability of the gravitational coupling constant not break any hope of making sense of the theory?

(3): What is the relation between Euclidean and Lorentzian signature in the absence of any Osterwalder-Schrader axioms to ensure that we can rotate from Euclidean space to Lorentzian space-time?

(4): What is the role of topology, keeping in mind that four-dimensional topologies cannot be classified?

One could hope that a non-perturbative definition of quantum gravity could help us in understanding the first three points. We propose a such a definition [19, 20, 21] here.

Not much is known about the question of summing over topologies. As long as we think in terms of continuum physics and write down the path integral it offers the possibility of summing over different manifold structures as well as integrating over inequivalent Riemannian structures of a given manifold:

$$Z = \sum_{\mathcal{M} \in \text{Top}} \int_{\mathcal{M}} \frac{\mathcal{D}g_{\mu\nu}}{\text{Vol(diff)}} e^{-S[g]}. \tag{5.1}$$

In two and three dimensions we do not have to worry about the meaning of *Top*, since there is equivalence between smooth manifolds and topological manifolds. In two dimensions the manifolds are uniquely characterized by their Euler characteristic $\chi$ and the summation over *Top* is simply a summation over $\chi$ or the genus $g = 1 - \chi/2$ of the surfaces. In spite of this simple prescription surprisingly little progress was made in defining the sum in (5.1) using continuum methods. The matrix model formulation of simplicial quantum gravity at least allowed us to address the question in a quantitative way by means of analytic continuations of the integral (3.8), but so far the results were ambiguous. If we move to three dimensions we encounter a slight classification problem in the sense that no simple parametrization of the various topologies exists. However, the problem gets completely out of control when we move to four dimensions. For four-dimensional manifolds there is no equivalence between smooth and topological structures. Topological manifolds exist which do not admit smooth structures and some topological manifolds admit infinitely many inequivalent smooth structures. If we insist on summing over all smooth structures $\sum_{Top}$ will be rather unwieldy. To complicate the operational meaning of $\sum_{Top}$ it should be added that four-dimensional manifolds are not algorithmic classifiable, i.e. no finite algorithm *in the sense of Turing* exists which allows us to decide if two arbitrary four dimensional manifolds are equivalent. On the other hand arguments (not known to us) might exist which dictate a restriction of the allowed class of manifolds. Since there seem to be fermions in the world one could argue that the manifold should be a spin manifold. If one makes the additional (rather arbitrary) restriction that it should be simply connected such a (smooth) manifold is characterized by its Euler number and its signature where these in four dimensions are given by:

$$\chi(M) = \frac{1}{128\pi^2} \int_M d^4\xi \sqrt{g}\, \mathcal{R}_{abcd}\mathcal{R}_{a'b'c'd'}\varepsilon^{aba'b'}\varepsilon^{cdc'd'} \tag{5.2}$$

$$\tau(M) = \frac{1}{96\pi^2} \int_M d^4\xi \sqrt{g}\, \mathcal{R}_{abcd}\mathcal{R}^{ab}{}_{c'd'}\varepsilon^{cdc'd'} \tag{5.3}$$

For simply connected spin manifolds the signature $\tau$ is a multiple of 16, while $\chi$ is an integer $\geq 2$. Thus eq. (5.3) seems a minor extension compared to two dimensions where the summation is over $\chi$, but the restriction to simply connected spin manifolds is not natural at this stage of a quantum theory of space time.

While these problems tend to discourage any attempt to make sense of the path integral of Euclidean quantum gravity, it is still our obligation to try to investigate if it is possible. Below we will argue that the use of dynamical triangulations, which work so well in two dimensions, allows us to discuss several of the above issues in some detail and might be a candidate for non-perturbative definition of quantum gravity even in higher dimensions.

## 5.2 Definition of simplicial quantum gravity for $d > 2$

Let us define simplicial quantum gravity as a generalization of the two-dimensional construction in two dimensions: In $d$ dimensions (where $d = 3$ or $d = 4$) we construct all closed (abstract) simplicial manifolds from $K$ $d$-dimensional simplexes. As in two dimensions we imagine that the lengths of the links in the simplexes are $a$ (which we take as 1 unless explicitly stated). For such a combinatorial or, equivalently, piecewise linear manifold we can apply Regge calculus and in this way assign a Riemannian metric to the manifold. By such an assignment we see that the discretization is able, for finite $K$, to assign a meaning to the sum and integral in (5.1)

$$\sum_{\mathcal{M} \in \text{Top}} \int_\mathcal{M} \frac{\mathcal{D}g_{\mu\nu}}{\text{Vol(diff)}} \to \sum_T \frac{1}{C_T}. \tag{5.4}$$

*The discretization has the same virtue as in two dimensions: In principle it allows a unified treatment of the summation over topologies and Riemannian structures.*

Few remarks should be said about the formula (5.4). First one could try as in two dimensions to make the gluing automatic, and in this way arrive at generalized matrix models [27]. Until now these models have not been as useful as in two-dimensions, the reason being that although the models allow a $1/N$ expansion, the two-dimensional interpretation as an expansion in topology is absent. Secondly, in light of the complicated relation between topology and diffeomorphism for four-dimensional manifolds one could be worried that similar problems arise when we compare combinatorial, i.e. piecewise linear, manifolds and smooth manifolds. However, for dimensions $d < 7$ we have equivalence between piecewise linear and smooth structures. Whatever subtleties might be involved in defining the sum on the lhs of eq. (5.4) it should be captured at the rhs of eq. (5.4). Of course eq. (5.4) itself is rather formal as it stands. The problem, which exists even in two dimensions, is that if no restriction is imposed on topologies the number of triangulations grows at least as fast as $K!$, $K$ being the number of triangles. No reasonable discretized action can kill this entropy factor and

$$Z = \sum_T e^{-S_T} \tag{5.5}$$

will be divergent. The naive expression (5.5) makes no sense. In two dimensions the way out was to fix topology. This bounds the number of triangulations exponentially and eq. (5.5) will be well defined. Only afterwards the topology was allowed to fluctuate and a summation attempted in the context of matrix integrals. In higher dimensions it is sensible, as a minimum, to try to define eq. (5.5) for a fixed topology. We need the following theorem

**Theorem:** The number of combinatorially equivalent $d$-dimensional manifolds is an exponentially bounded function of the number of $d$ dimensional simplexes.

We call two simplicial complexes combinatorially equivalent if they have a common subdivision and when we talk about equivalence classes of piecewise linear manifolds we have in mind combinatorial equivalence. While the theorem has been known in two dimensions for many years and it has been rather convincingly established numerically in three dimensions [22], there have recently been some controversies concerning its validity in four dimensions [23], but now there appeared a proof which is valid in any dimension and for any fixed topology [24].

For $d \geq 4$ manifolds exist which are not algorithmic recognizable. As a curiosity we can mention that for such manifolds their number is not algorithmic calculable. This does not mean that the number cannot be exponentially bounded, only that there is no finite algorithm which allows us to calculate the exact number.

After these general remarks it is natural as a first exploratory step to fix the topology of our four-dimensional manifold to be the simplest possible, that of $S^4$. The Einstein-Hilbert action for a simplicial manifold can be calculated by Regge calculus. However, we do not need the full machinery in our case where the simplexes are identical and all link length equal. The Regge version of the Einstein action is the sum over deficit angles of the $d - 2$ dimensional sub-simplexes times their $d - 2$ dimensional volume. In our case the deficit angle associated with a $d - 2$ dimensional sub-simplex $n_{d-2}$ is $2\pi - c \cdot o(n_{d-2})$, where $o(n_{d-2})$ is the *order* of $n_{d-2}$, i.e. the number of $d$ dimensional simplexes of which $n_{d-2}$ is a sub-simplex, and $c$ is a constant. It follows that

$$\left\{\int d^d\xi \sqrt{g} \mathcal{R}\right\}_{\text{Regge}} \propto \sum_{n_{d-2}} (2\pi - c \cdot o(n_{d-2})). \tag{5.6}$$

If we note that the number of $d - 2$-dimensional sub-simplexes in a $d$-dimensional simplex is $d(d+1)/2$ we can write:

$$S[\Lambda, G] = \int d^d\xi \sqrt{g}\left(\Lambda - \frac{1}{16\pi G}\mathcal{R}\right) \to S_T[k_d, k_{d-2}] = k_d N_d(T) - k_{d-2} N_{d-2}(T), \tag{5.7}$$

where the piecewise linear manifold is defined by the triangulation $T$ and $N_d(T)$ and $N_{d-2}(T)$ denote the number of $d$- and $d - 2$-dimensional simplexes in the triangulation $T$.

We can view $1/k_{d-2}$ as a bare gravitational coupling constant. At first sight the action might seem much too simple to have anything to do with gravity. Our point of view will be the opposite: *The fact that the action is so simple reflects the beauty and simplicity of quantum gravity, and hopefully this simplicity will be reflected in the solution of the theory.*

Our final prescription will be:

$$Z[\Lambda, G] = \int_{S^4} \frac{\mathcal{D}g_{\mu\nu}}{\text{Vol}(\text{diff})} e^{-S[g;\Lambda,G]} \to Z[k_{d-2}, k_d] = \sum_{T \in S^d} e^{-S_T[k_{d-2},k_d]}. \quad (5.8)$$

We will be interested in $d = 4$ and sometimes in $d = 3$, in which cases we get

$$Z[k_1, k_3] = \sum_{T \in S^3} e^{-k_3 N_3(T) + k_1 N_1}. \quad (5.9)$$

$$Z[k_2, k_4] = \sum_{T \in S^4} e^{-k_4 N_4(T) + k_2 N_2}. \quad (5.10)$$

For $d = 3, 4$ there are only two independent coupling constants as long as we only want an action which depends on global quantities like $N_i$. It follows from the so called Dehn-Sommerville relations which express that the $d$-dimensional simplicial neighborhood of any sub-simplex is homomorphic to a $d$-simplex:

$$N_i = \sum_{k=1}^{d} (-1)^{k+d} \binom{k+1}{i+1} N_k. \quad (5.11)$$

If we include higher derivative terms in the action we will certainly loose the simplicity of eqs. (5.9) and (5.10). The higher derivative terms will contain explicitly the order of the sub-simplexes which carry the curvature. We will not discuss the lattice implementation of these.

Let us now discuss the phase diagram of the discretized theory. Assume $d = 4$. Since it is easy to prove that $N_2(T) \leq c \cdot N_4(T)$, where $c$ is some constant, the conjecture above implies that for each $k_2$, a $k_4^c(k_2)$ exists such that the lhs of (5.10) for a given $k_2$ is well-defined for $k_4 > k_4^c(k_2)$ and divergent for $k_4 < k_4^c(k_2)$. A potential continuum limit should be taken as $k_4 \to k_4^c$ from above. The corresponding phase diagram is shown in fig. 13. If for a fixed $k_2$ approaching $k_4^c$ we shall be probing the infinite volume limit of the discretized system. It does not imply that there necessarily will be a continuum limit. Rather we should view the system as a lattice system where the infinite volume limit is taken. For some specific values of the bare couplings critical points might exist where a correlation length diverges and where a continuum limit exists. Such a point is tentatively indicated at the figure. Approaching this point in a specific way will then define the renormalized cosmological constant and the renormalized gravitational constant. Since we are in unchartered territory one should be open-minded for other possibilities, e.g. the possibility that a whole range of $k_2$'s might correspond to topological gravity where the metric, and correspondingly concepts like volume and divergent distances, play no role.

### 5.3 Observables

It is possible to define the same critical exponents for higher dimensional simplicial quantum gravity as in two-dimensional quantum gravity.

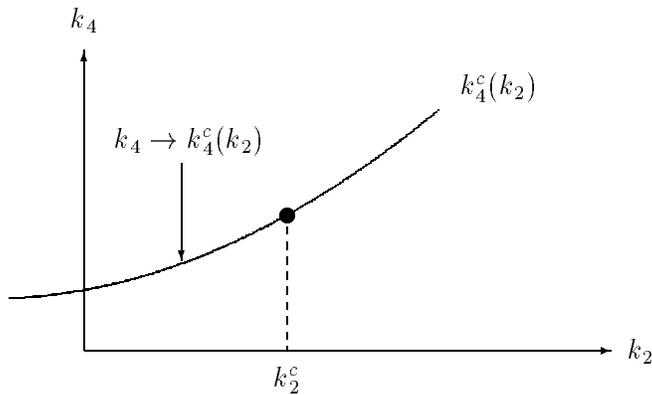

Figure 13: A hypothetical phase diagram for four-dimensional gravity.

First we can define an entropy or susceptibility exponent $\gamma$. In the following it will always be assumed that the topology is spherical, i.e. the combinatorial manifolds are combinatorially equivalent to the boundary of a 5-simplex. According to the theorem mentioned above the number $\mathcal{N}(N_4)$ of triangulations which can be constructed from $N_4$ 4-simplexes is exponentially bounded. Let us now fix $k_2$. As already argued there will be a critical point $k_4(k_2)$. This implies that

$$Z(k_2, N_4) \equiv \sum_{T \in S^4(N_4)} e^{k_2 N_2} \sim e^{k_4^c(k_2) N_4} f_{k_2}(N_4) \tag{5.12}$$

where $f_{k_2}(N_4)$ stands for subleading corrections. *If* the subleading correction is power-like we define $\gamma(k_2)$ by

$$Z(k_2, N_4) \sim N_4^{\gamma(k_2)-3} e^{\kappa_4^c N_4}(1 + O(1/N)). \tag{5.13}$$

However, it is not clear that we have such a behavior. It is possible to imagine an asymptotic behavior like:

$$Z(k_2, N_4) \sim e^{\kappa_4^c N_4 - c(k_2) N_4^\alpha(k_2) + \cdots}(1 + O(1/N)) \tag{5.14}$$

where $0 < \alpha < 1$. In this case the exponential correction given by $\alpha$ will always dominate over the power-like correction determined by $\gamma(k_2)$. There are strong indications that there are several regions with different asymptotic behavior, depending on the value of $k_2$. This will be discussed below.

Apart from the entropy- or susceptibility exponent $\gamma$ we can introduce the critical exponents $\nu$ and $\eta$ already discussed in detail in the context of two-dimensional quantum gravity, where they were determined from the properties of the two-point function. Precisely the same construction as in two dimensions can be carried out in higher dimensions and the definitions of Hausdorff dimension etc. is basically the same.

## 5.4 Results of numerical simulations

Presently there has not been much progress in analyzing (5.10) by analytic methods except for $d = 2$. However, the action is well suited for the use of Monte Carlo simulations and the results to be discussed later have been obtained by such simulations. For details about

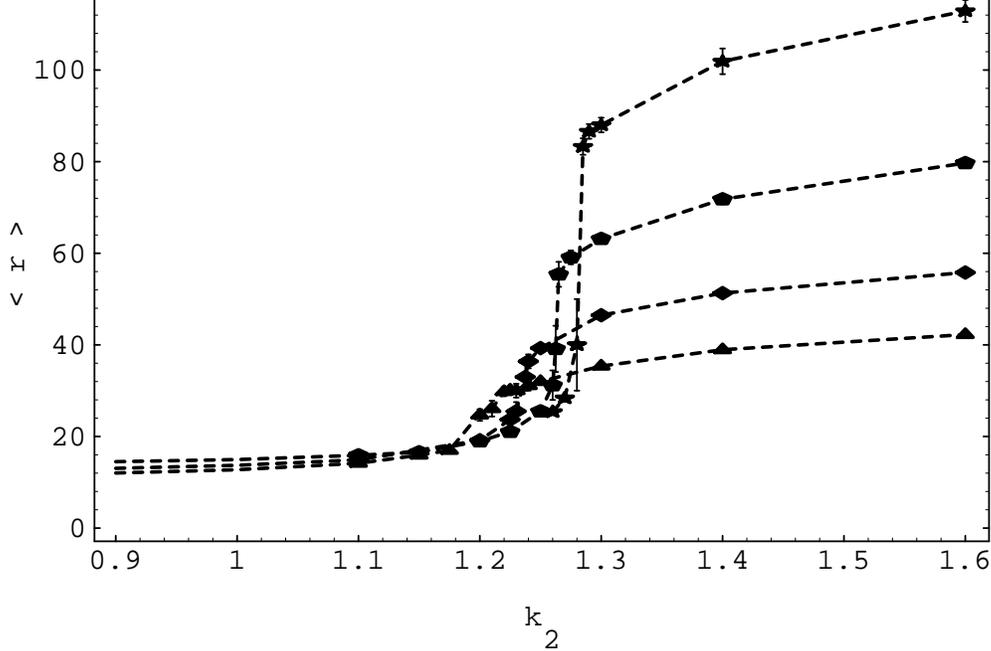

Figure 14: The average radius of the universes of sizes 9000 (triangles), 16000 (squares), 32000 (pentagons) and 64.000 (stars) versus $k_2$.

the most recent simulations we refer to [26]. Here let us only discuss the results. They can be summarized as follows [20, 21, 25, 26, 27]

(1): There seem to be two different regions, as a function of the bare *inverse* gravitational coupling constant $k_2$: For small or negative values of $k_2$ the typical quantum universe will be very crumpled, with almost no extension and a very large, if not infinite Hausdorff dimension, while the universes for large values of $k_2$ will be elongated with a Hausdorff dimension as small as two. In fig. 14 we have shown the average radius for universes of size 9000, 16000, 32000 and 64000 4-simplexes as a function of $k_2$. The two phases are separated by a phase transition which is of order two or higher. *At* the transition point, $k_2^c$, the Hausdorff dimension might be finite (the precise value is not well determined, but it could be close to four).

(2): The same results are valid for three dimensional simplicial quantum gravity except that the phase transition seems to be of *first order*, rather than of higher order [27].

From fig. 14 it is seen that the change between the elongated region and the crumpled region becomes increasing visible as the size of the system increases. In addition the critical point seems to move to higher values of $k_2$. We observe a so called pseudo critical point $k_2^c(N_4)$. The indication of convergence to a limiting value $k_2^c$ as the volume $N_4 \to \infty$ is shown in fig. 15 and we conclude that we have a genuine phase transition. From the general theory of finite size scaling the pseudo critical point $k_2^c(N_4)$ of a first order phase transition scales to the limiting value $k_2^c$ as $1/N_4$. Since we observe a scaling like $1/\sqrt{N_4}$ we conclude that the transition most likely is a higher order transition. In three dimensional simplicial quantum gravity it is difficult to perform this kind of measurement since we observe pronounce hysteresis around the transition point. This is a typical sign of a *first order* transition.

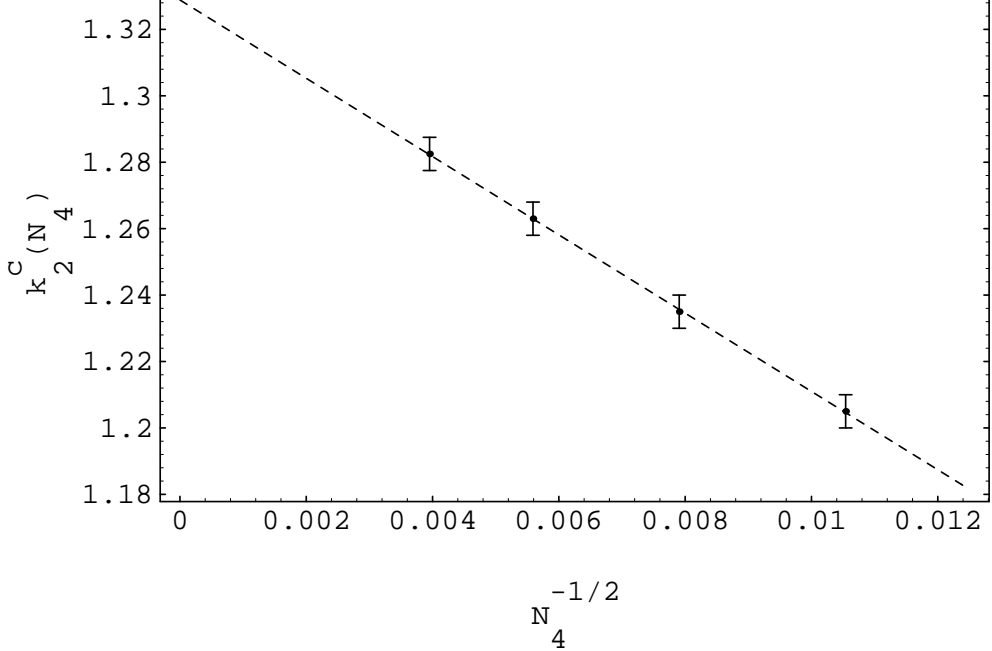

Figure 15: The pseudo critical point $k_2(N_4)$.

Let us first discuss the measurements in the elongated phase, i.e. for $k_2 > k_2^c$. An ideal quantity to measure in the computer simulations is the number of four-simplexes $\langle n(r) \rangle_{N_4}$ within a spherical shell of thickness 1, a geodesic distance $r$ from a marked four simplex, the average taken over all spherical triangulations with one marked four-simplex. It will depend on the coupling constant $k_2$ and it is related to the two-point function precisely as in two dimensions:

$$G(r, N_4) \sim N_4^{\gamma(k_2)-2} e^{k_4^c(k_2) N_4} \langle n(r) \rangle_{N_4}, \qquad (5.15)$$

provided the subleading corrections are power-like. In fig. 16 we show the measured distribution $\langle n(r)/N_4 \rangle_{N_4}$ just above the transition to the elongated phase and a corresponding fit to $r \exp(-r^2/N_4)$. We see a very good agreement and it becomes even better if we move further into the elongated phase. Now recall the general scaling relations derived in the context of two-dimensional quantum gravity, but valid in any dimension if a finite Hausdorff dimension exists:

$$\langle n(r) \rangle_{N_4} \sim r^{d_h-1} \quad \text{for} \quad 1 \ll r \ll N_4^{1/d_h}, \qquad (5.16)$$

$$\langle n(r) \rangle_{N_4} \sim r^\alpha e^{-c(r^{d_h}/N_4)^{1/(d_h-1)}} \quad \text{for} \quad N_4^{1/d_h} < r < N_4, \qquad (5.17)$$

we conclude that $d_h = 2$ (and $\alpha = 1$). In addition we can measure the critical exponent $\gamma$ very conveniently by baby universe counting. Again the arguments are identical to ones presented in the two-dimensional case. The result is shown in fig. 17 and it is natural from the figure to conjecture that $\gamma = 1/2$ as long as we are in the elongated phase. From (5.15) and $\langle n(r) \rangle_{N_4}$ we know $G(r, N_4)$ and we can construct the two-point function $G_{k_4}(r; k_2))$ by Laplace transform (as in the two-dimensional case):

$$G_{k_4}(r; k_2) = \sum_{N_4=1}^\infty N^{-3/2} e^{-\Delta(k_4) N_4} r e^{-cr^2/N_4} \sim e^{-\tilde{c} r \sqrt{\Delta k_4}}, \qquad (5.18)$$

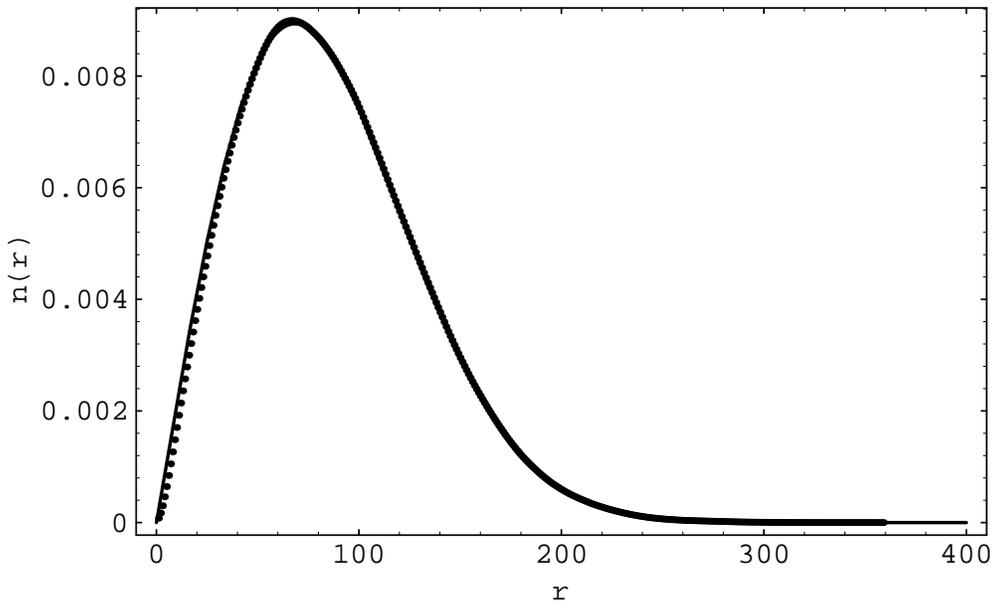

Figure 16: The measured distribution (dots, the error bars smaller than the dots), and the fit (the curve) using the functional form $r\exp(-r^2/N_4)$.

where $\Delta(k_4) = k_4 - k_4^c(k_2)$ is assumed to be small. This function is precisely the two-point function of the so called branched polymers, which are known to have internal Hausdorff dimension $d_h = 2$. We conclude that the numerical simulations provide convincing evidence that the elongated phase of simplicial quantum gravity corresponds to a well known statistical theory, the one of branched polymers. *The tendency to create baby universes is so pronounced in this phase that the geometry degenerates to the generic lowest dimensional fractal structure possible, i.e. that of branched polymers.*

When we lower the value of $k_2$ and move below the critical point $k_2^c$ the fractal structure of our ensemble of piecewise linear manifolds changes drastically. A glance on fig. 14 shows that the average radius hardly changes with the volume. This is an indication that the Hausdorff dimension is large or maybe infinite. If we move deep into this phase the average curvature is negative and in addition there are only few baby universes and they are small. This could lead to the idea that we entered a phase with "smooth" manifolds of negative curvature. For such manifolds one would expect that the volume of geodesic balls of radius $r$ would grow exponentially with the radius, which is what we observe. Clearly this is a "fake" infinite Hausdorff dimension and indeed we should observe the dimension $d_h = 4$ in the sense that $\langle n(r)\rangle_{N_4} \sim r^3$ for small geodesic distances. A closer look at "typical" members of the computer generated manifolds indicates that they cannot be considered as "smooth". Rather they have a few vertices of very high orders which connect to almost any other vertex in the manifold and in such a situation it is not surprising that the linear extension will be small. In addition we have not been able to fit to any sensible power dependence $r^{d_H-1}$ for small $r$. A plot of $\log\langle n(r)\rangle_{N_4}$ shows indeed a linearly growing function of $r$ up to some $r_0(N_4)$ which is not much different from the average value $\langle r\rangle_{N_4}$.

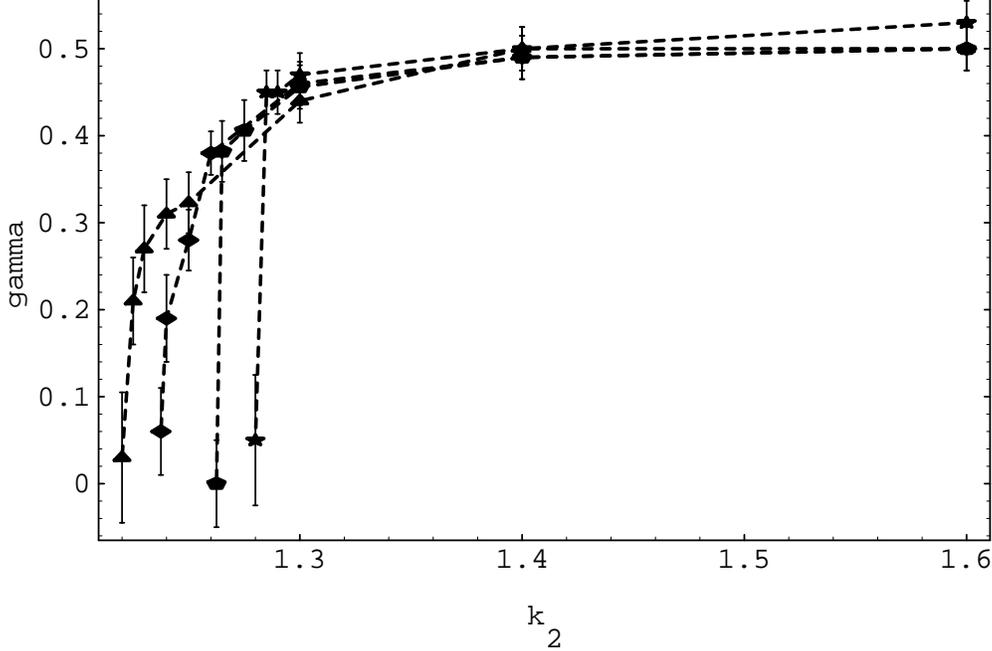

Figure 17: The measure $\gamma$ in the elongated phase for various size lattices ($N_4 = 9000$, 16000, 32000 and 64000).

A reasonable fit to $\langle r \rangle_{N_4}$ is

$$\langle r \rangle_{N_4} = a(k_2) + b(k_2) \log N_4. \tag{5.19}$$

This again gives some support to the idea that the Hausdorff dimension is infinite in this phase since it appears as a limit of $\langle r \rangle_{N_4} \sim N^{1/d_H}$ for $d_H \to \infty$. Finally the extrapolation of (5.17) to infinite Hausdorff dimension indicates that in such a case one should observe

$$\langle n(r) \rangle_{N_4} \sim c_1 \, e^{-m_1(k_2) r} \tag{5.20}$$

down to quite small distances $r \gg N_4^{1/d_h}$, $d_h \to \infty$, which naturally is replaced by $r > a_1 \log N_4 + a_2$. This is indeed what we measure.

The observation that $\langle n(r) \rangle_{N_4}$ grows exponentially from $r \approx 6$ out to $r \approx r_0$ and then falls off exponentially indicates that we deal with an infinite Hausdorff dimension at all distances and it is easy to get a quite good "phenomenological" fit to $\langle n(r) \rangle_{N_4}$ which incorporates both these features by choosing e.g.:

$$\langle n(r) \rangle_{N_4} \sim \exp\left(-m_1(k_2) r - c_2 e^{-m_2(k_2) r}\right). \tag{5.21}$$

It will grow like $e^{(c_2 m_2 - m_1) r}$ for small distances and fall off like

$$\langle n(r) \rangle_{N_4} \sim c_1 \, e^{-m_1(k_2) r} - c_2 \, e^{-(m_1(k_2) + m_2(k_2)) r} + \cdots \tag{5.22}$$

for large distances, while a $N_4$ dependence in the coefficient $c_2$ would explain the observed $N_4$ dependence of $r_0$. The data and a fit of the form (5.21) are shown in fig. 18 for $k_2 = 1.26$ and $N_4 = 64000$, i.e. right below the transition to the crumpled phase, where the fit is worst. But even so close to critical point (5.21) works quite well over the whole range of $r > 6$. It should be mentioned that the coefficient in front of the second exponential in eq. (5.22) is negative. This implies that the term cannot be given the interpretation as an additional heavier mass excitation. However, just looking at the long distance tail the distribution $\langle n(r) \rangle_{N_4}$ allows us to determine $m_1$, $m_2$ and $c_2$ from (5.22). On the other hand we can determine $c_2 m_2 - m_1$ from the short distance exponential growth alone and find good agreement. This indicates that long and short distance behavior are intertwined in the case of infinite Hausdorff dimension, as they are in the case of finite Hausdorff dimension, where $d_H$ appears both in the short distance and long distance expression for $\langle n(r) \rangle_{N_4}$ (see e.g. (5.16)-(5.17)).

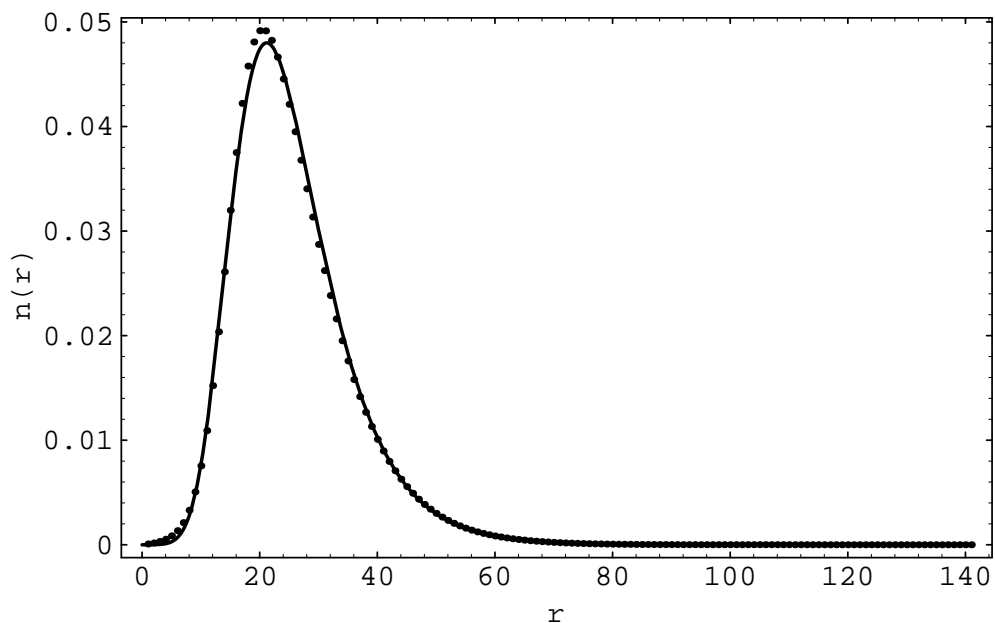

Figure 18: exponential fit (curve) of the form (5.21) to the measured $n(r; N_4)$ (dots, error bars less than dot-size) in the crumpled phase ($N_4 = 64000$, $k_2 = 1.26$).

*We conjecture that the internal Hausdorff dimension is infinite for $k_2 < k_2^c$.*

In accordance which this conjecture we have in the elongated phase no "mass" term $m(k_4) = (k_4 - k_4^c(k_2))^\nu$ which scales to zero as we approach $k_4^c(k_2)$. The exponential coefficient $m_1(k_2)$ is finite in the infinite volume limit $k_4 \to k_4^c(k_2)$. However, it is most interesting that $m_1(k_2)$ scales to zero as $k_2 \to k_2^c$, the phase transition point between the crumpled and the elongated phase. *This gives a strong indication that the system at the transition might have a finite Hausdorff dimension, which could very well be larger that the generic $d_h = 2$ found in the elongated phase.* In fig. 19 we have shown the scaling of the mass $m_1(k_2)$ as a function of $k_2$.

The above mentioned numerical "experiments"[9] suggest the following scenario: The typical quantum universe, determined without any Einstein action (i.e. $k_2 = 0$) has (almost) no extension. Its Hausdorff dimension might be infinite and internal distances between points are always "at the Planck scale". By that we simply mean that no consistent scaling can be found which will be compatible with finite continuum volume and finite Hausdorff dimension. For a finite value of the bare gravitational coupling constant there is a phase transition (second or higher order) to a phase with a completely different geometry with pronounced fractal structures. It is tempting to view the transition between the two kind of geometries as a transition where excitations related to the conformal mode are liberated, since large $k_2$ is a region which formally corresponds to small values of the gravitational coupling constant. Right at the transition it seems as if we have the chance to encounter genuine extended structures with a finite Hausdorff dimension. Maybe the fact that the transition between the two types of geometry is of second (or higher) order can be used as the starting point for a non-perturbative definition of quantum gravity.

---

[9]See [26] for a detailed discussion of the most recent results.

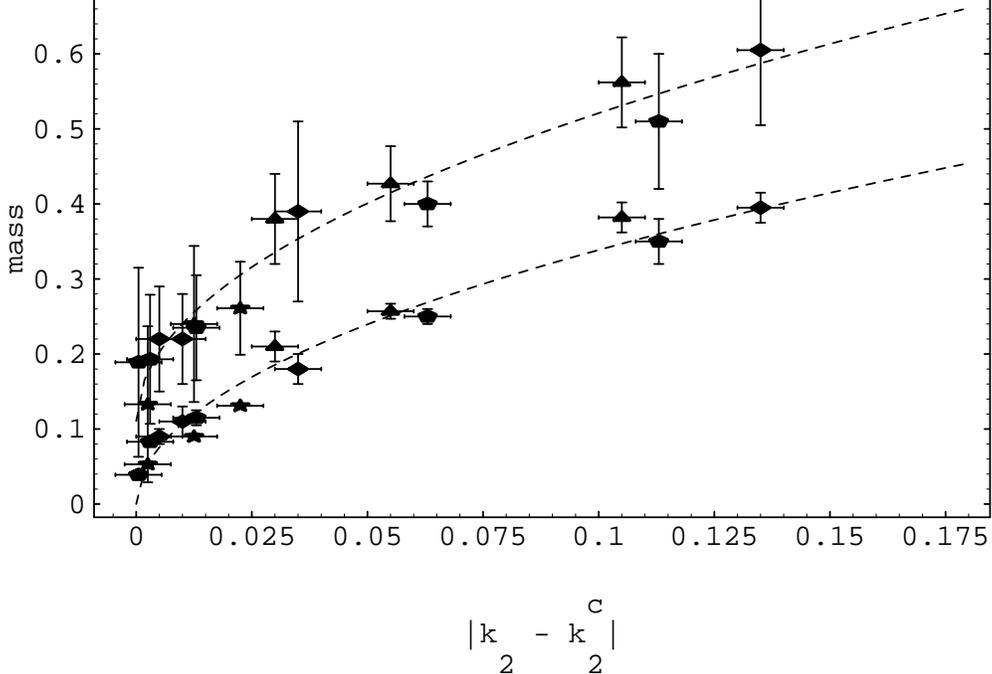

Figure 19: The behavior of the masses $m_1(k_2)$ and $m_1(k_2) + m_2(k_2)$ from (5.21) in the crumpled phase as a function of $k_2^c - k_2$.

Of course it is crucial to be able to perform high statistics simulations *at* the critical point in order to investigate this possibility in greater detail.

## 6  Discussion

We have shown how it is possible to discretize reparametrization invariant theories and apply with success the methods known from theory of critical phenomena. In two-dimensional gravity it was possible to solve the theory and calculate the Hartle-Hawking wave functional and reparametrization invariant observables like the volume-volume correlator as a function of the geodesic distance. These observables are not easily calculable using a continuum framework like Liouville theory, not to mention canonical quantization.

In three- and four-dimensional gravity we also obtained a well defined theory with specific scaling relations and it was possible to determine these reliably by numerical methods in part of the parameter space, the parameters being the bare gravitational and the bare cosmological coupling constants. An interesting phase transition point was found in four dimensions and it is a potential candidate for the point in coupling constant space where a non-perturbative definition of quantum gravity might lead to a physically acceptable theory.

It is interesting to compare the results obtained with theoretical suggestions. The simplest comparison is made with so called "conformal gravity". Presently there are two versions which are quite different: The first one is based on adding higher derivative terms to the Lagrangian and taking the coupling constant in front of the square of the Weil tensor to infinity [28], while the other approach appeals to the scale anomaly as the factor which dictates the important physics [29]. Both theories have a non-trivial fixed point. In the first case it is an ultraviolet fixed point, in the second case an infrared fixed point, so physics is quite different for the two theories. They both give definite scaling predictions for the partition function and it seems as if there is surprisingly good agreement between the predictions coming from the theory with an infrared fixed point and simplicial gravity at $k_2^c$. However, the comparison is not unambiguous and it is too early to make definite statements. This work is in progress.

At the moment we let loose topology we also loose control of the theory. In two dimensions the use of dynamical triangulations led to a number of very interesting attempts to perform the summation over topologies by means of analytic continuation of the matrix integrals which can be used to represent the partition function. It is a most important task to understand if it is also possible in higher dimensions to address in a quantitative way the summation over topologies. Dynamical triangulations will hopefully help us to address the question and force us to ask the right questions, since the formalism is so simple, that it is impossible to hide in mathematical abstractions.

# References

[1] T. Regge, Nuovo Cimento 19 (1961) 558.

[2] F. David, Nucl.Phys. B257 (1985) 45; B257 (1985) 543.

[3] J. Ambjørn, B. Durhuus and J. Fröhlich, Nucl.Phys. B257 (1985) 433.

[4] V.A. Kazakov, I. Kostov and A.A. Migdal, Phys.Lett. B157 (1985) 295.

[5] W.T. Tutte, Can.J.Math. 14 (1962) 21.
F.A. Bender and E.R. Canfield, *The asymptotic number of rooted maps on a surface*, 1986.

[6] S. Jain and S.D. Mathur, Phys.Lett. B286 (1992) 239.

[7] J. Ambjørn and B. Durhuus and T. Jonsson, Mod.Phys.Lett. A6 (1991) 1133.

[8] J. Ambjørn and Y. Watabiki, *Scaling in quantum gravity* NBI-HE-95-01, hep-th/9501049.

[9] J. Ambjørn and J. Jurkiewicz,

[10] J. Ambjørn, B. Durhuus and T. Jonsson, Phys.Lett. B244 (1990) 403.

[11] J. Ambjørn, B. Durhuus J. Fröhlich and P. Orland, B270 (1986) 457; B275 (1986) 161.

[12] J. Ambjørn, J. Jurkiewicz and Yu.M. Makeenko, Phys.Lett. B251 (1990) 517.

[13] J. Ambjørn, L. Chekhov, C.F Kristjansen and Yu. Makeenko, Nucl.Phys. B404 (1993) 127; Modern Physics Letters A7 (1992) 3187-3202.

[14] S. Wadia, Phys.Rev. D24 (1981) 970.
A.A. Migdal, Phys.Rep. 102 (1983) 199.
F. David, Mod.Phys.Lett. A5 (1990) 1019.

[15] V.A. Kazakov, Mod.Phys.Lett. A4 (1989) 2125.

[16] T. Morris, Nucl.Phys. B356 (1991) 703.
S. Dalley, C. Johnson, T. Morris, Nucl.Phys. B368 (1992) 625.

[17] H. Kawai, N. Kawamoto, T. Mogami and Y. Watabiki Phys.Lett.B306 (1993) 19.
N. Kawamoto, Y. Saeki, and Y. Watabiki, in preparation;
N. Kawamoto, INS-Rep.972;
Y. Watabiki, Prog.Theor.Phys.Suppl. No.114 (1993) 1.

[18] Y. Watabiki, *Construction of Non-critical String Field Theory by Transfer Matrix Formalism in Dynamical Triangulation*, INS-Rep.1017, to be published in Nucl. Phys. B.

[19] D. Weingarten, Nucl.Phys. B210 (1982) 229.

[20] J. Ambjørn and J. Jurkiewicz, Phys.Lett. B278 (1992) 50.

[21] M.E. Agishtein and A.A. Migdal, Mod. Phys. Lett. A7 (1992) 1039.

[22] J. Ambjørn and S. Varsted, Phys.Lett. B266 (1991) 285.
S. Catterall, J. Kogut and R. Renken, *Entropy and the approach to the thermodynamic limit in three-dimensional simplicial gravity.*, CERN-TH-7404-94, hep-lat-9408015.

[23] J. Ambjørn and J. Jurkiewicz, Phys.Lett. B335 (1994) 355.
B. Bruegmann and E. Marinari, *More on the exponential bound of four dimensional simplicial quantum gravity* MPI-PhT/94-72, hep-th/9411060
S. Catterall, J. Kogut and R. Renken, Phys. Rev. Lett. 72 (1994) 4062.

[24] C.Bartocci, U.Bruzzo, M.Carfora and A.Marzuoli, *Entropy of Random Coverings and 4-d Quantum Gravity*, SISSA Ref. 97/94/FM.

[25] J. Ambjørn, J. Jurkiewicz and C.F. Kristjansen, Nucl.Phys. B393 (1993) 601.
M.E. Agishtein and A.A. Migdal, Nucl.Phys. B385 (1992) 395.
J. Ambjørn, S. Jain, J. Jurkiewicz and C.F. Kristjansen, Phys.Lett. B305 (1993) 208.
J. Ambjørn, Z. Burda, J. Jurkiewicz and C.F. Kristjansen, Phys.Rev. D48 (1993) 3695.

    B. Bruegmann, Phys. Rev. D47 (1993) 3330.
    B. Bruegmann and E. Marinari, Phys. Rev. Lett. 70 (1993) 1908.
    B.V. De Bakker and J. Smit, Phys.Lett. B334 (1994) 304.
    B.V. De Bakker and J. Smit, *Curvature and scaling in 4D dynamical triangulation*, preprint ITFA-94-23. hep-lat/9407014.
    S. Catterall, J. Kogut and R. Renken, Phys.Lett. B328 (1994) 277.

[26] J. Ambjørn and J. Jurkiewicz, *Scaling in four dimensional quantum gravity*, NBI-HE-95-05, hep-th/9503006.

[27] M.E. Agishtein and A.A. Migdal, Mod. Phys. Lett. A6 (1991) 1863.
    B. Boulatov and A. Krzywicki, Mod.Phys.Lett A6 (1991) 3005.
    J. Ambjørn and S. Varsted, Nucl.Phys. B373 (1992) 557.
    J. Ambjørn, D.V. Boulatov, A. Krzywicki and S. Varsted, phys.Lett. B276 (1992) 432.

[28] E.S. Fradkin and A.A. Tseytlin, Nucl. Phys B201 (1982) 469.
    C. Schmidhuber Nucl.Phys.B390 (1993) 188.

[29] I. Antoniadis and E. Mottola, Phys.Rev. D45 (1992) 2013.
    I. Antoniadis, P.O. Mazur and E. Mottola, Nucl.Phys. B388 (1992) 627.
    I. Antoniadis, P.O. Mazur and E. Mottola, Phys.Lett. B323 (1994) 284.